\documentclass[preprint,aps,showpacs]{revtex4}
\usepackage{amssymb}
\usepackage{bm}
\usepackage{array}
\usepackage{graphicx}
\usepackage{caption2}
\usepackage{amsmath}
\usepackage{subfigure}
\usepackage{natbib}

\begin{document}
\title{Multiatom and resonant interaction scheme for quantum state
  transfer and logical gates between two remote cavities via an
  optical fiber} 
\author{Zhang-qi Yin}
\author{Fu-li Li}
\email[Email: ]{flli@mail.xjtu.edu.cn}
 \affiliation{Department of
Applied Physics, Xi'an Jiaotong University, Xi'an 710049, China}
\begin{abstract}
A system consisting of two single-mode cavities spatially separated
and connected by an optical fiber and multiple two-level atoms trapped
in the cavities is considered. If the atoms resonantly and
collectively interact with the local cavity fields but there is no
direct interaction between the atoms, we show that an ideal quantum
state transfer and highly reliable quantum swap, entangling, and
controlled-Z gates can be deterministically realized between the
distant cavities. We find that the operation of state transfer and
swap, entangling, and controlled-Z gates can be greatly speeded up as
number of the atoms in the cavities increases. We also notice that the
effects of spontaneous emission of atoms and photon leakage out of
cavity on the quantum processes can also be greatly diminished in the
multiatom case. 
\end{abstract}

\maketitle

\section{Introduction}
Quantum computers implement computational tasks on the basis of the
fundamental quantum principle. Shor \cite{Sho94} showed that a
quantum computer could complete the factorization of large composite
numbers into primes in only polynomial time, which is the basis of
the security of many classical key cryptosystems. Grover
\cite{Gro97} discovered a quantum algorithm that searches a number
from a disordered database polynomially faster than any classical
algorithms. The progress has greatly stimulated much interest in
quantum computer and quantum computation.

For building a quantum computer that could be used to solve some
practical problems, a large number of qubits such as either trapped
atoms or ions must be assembled together and manipulated according
to certain orders \cite{CZ00}. In order to work in this way,
\emph{distributed quantum computing} is introduced \cite{CE+99}.
Distributed quantum computing is an architecture in which a quantum
computer is thought as a network of spatially separated local
processors that contain only a few qubits and  are connected via
quantum transmission lines \cite{PK+03}. One of the key problems in
the realization of distributed quantum computing is how to implement
two-qubit quantum gates among local processors since a quantum
computer can be built by assembling two-quantum-bit logic gates
\cite{DiV95,BBC+95}. Controlled-phase gates between atoms trapped in
distant optical cavities have been recently
 proposed \cite{DK04, LBK05, BK05, CL05, XL+05}. Several schemes
which are based on cavity QED systems have also been proposed to
implement quantum communication or engineer entanglement between
atoms trapped in two distant optical cavities
\cite{CZKM97,Pel97,EK+99,CP+03,
 BP+03,DK03,MB04}. In some of these
schemes \cite{CZKM97,Pel97,EK+99,CP+03}, two spatially separated
cavities are directly connected to each another via quantum
channels. In order to realize quantum logical gates, accurately
tailored sequences of controlling pulses or adiabatic processes are
involved and considerable local operations are required. In other
schemes \cite{BP+03,DK03,MB04}, the detection of leaking photons is
involved. The quantum logical gates are realized only in a
probabilistic way and the success probability is highly dependent on
the efficiency of photon detectors. Therefore, it is highly desired
that deterministic quantum gates between two separated subsystems
can be implemented in coherent evolutions of the entire system.  In
the recent paper \cite{SM+06}, \citeauthor{SM+06} investigated the
possibility of realizing deterministic swap and controlled-phase
gates between two-level atoms trapped in separate optical cavities,
through a coherent resonant coupling mediated by an optical fiber.
In the scheme, each of the cavities contains a single two-level
atom, and the only local control required is the synchronized
switching on and off of the atom-field interaction in the distant
cavities, achievable through simple control pulses.

In practical situations, various decoherence processes such as
spontaneous emission of atoms and photon leakage out of cavities are
inevitable. In order to diminish the effects of dissipation
processes on quantum information processing, the operation time of
quantum gates must be much shorter than the characteristic time of
various relaxations. In the present study, we consider the scheme,
similar to that proposed by \citeauthor{SM+06} \cite{SM+06},
however, multi two-level atoms are trapped in each of the cavities
and a qubit is encoded in zero- and single-excitation Dicke states
of the atoms. We find that perfect quantum-state transfer, and
quantum swap, entangling and controlled-Z gates between the qubits
can be realized, and moreover the operation time of these quantum
processes is proportional to $1/\sqrt{N}$ where $N$ is number of the
atoms trapped in each of the cavities. Therefore,  the quantum
processes under consideration can be greatly speeded up and the
effects of spontaneous emission and photon leakage can be depressed
if the number of atoms is large. We also find that highly reliable
controlled-Z gate can be realized in the resonant interaction with
much shorter operation time than that in the non-resonant case which
was considered in \cite{SM+06}.

The paper is organized as follows. In section II, we introduce the
model under consideration. In Sections III, IV, V and VI, quantum
state transfer, and swap, entangling and controlled-Z gates are
investigated, respectively. In Section VII, the influence of atomic
spontaneous emission and photon leakage out of the cavities and
fiber on the quantum state transfer, and the swap, entangling and
controlled-Z gates is investigated. In section VIII, a summary is
given.

\section{Model}
\begin{figure}[htbp]
  \centering
    \includegraphics[width=16cm]{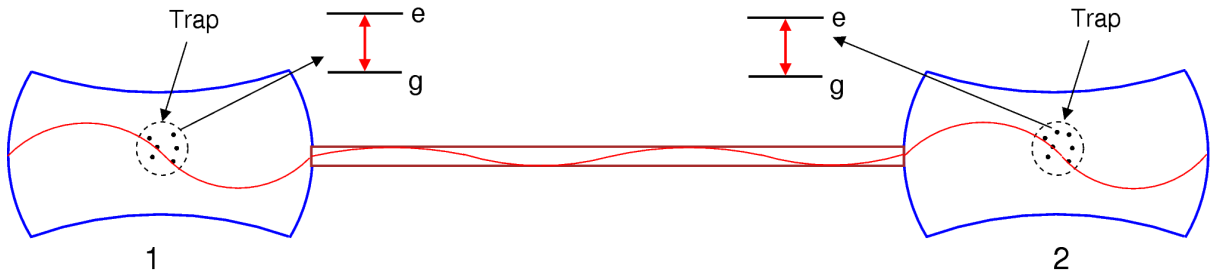}
  \caption{Experimental setup.}
  \label{fig:setup}
\end{figure}

As shown in Fig. \ref{fig:setup}, multi two-level atoms are trapped
in two distant single-mode optical cavities, which are connected by
an optical fiber. The atoms resonantly interact with the local
cavity fields. We assume that the size of the space occupied by the
atoms in each of the cavities is much smaller than the wavelength of
the cavity field. Then, all the atoms in each of cavities ``see''
the same field. On the other hand, the atoms in the same cavity are
so separated that they have no direct interaction each other.
The collectively raising and lowering operators for the atoms in
 cavity $j(=1,2)$ are defined as
\begin{equation}
\label{eq:J}
J^{\pm}_j = \sum_{i=1}^{N_j}\sigma^{\pm}_{i} (j),
\end{equation}
where $N_j$ is the number of atoms in cavity $j$, $\sigma^-_i(j) =
|g_i\rangle\langle e_i|$ and $\sigma^+_i(j) =(\sigma^-_i)^{\dagger}$
for atom $i$ in cavity $j$ with the ground state $\vert g\rangle$
and the excited state $\vert e\rangle$. In rotating wave
approximation, the interaction Hamiltonian of the atom-field system
can be written as
\begin{equation}
 H_{AF}= \sum_{j=1}^2 \Big(g_j J^-_j a^{\dagger}_j + \mathrm{h.c.}\Big),
\end{equation}
where  $a^{\dagger}_j$ is the creation operator for photons in the
mode of cavity $j$, and $g_j$ is the coupling constant between the
mode of cavity $j$  and the trapped atoms.

The coupling between the cavity fields and the fiber modes may be
modeled by the interaction Hamiltonian \cite{Pel97}
\begin{equation}
  \label{eq:HIF}
  H_{IF} = \sum_{j=1}^{\infty} \nu_j \Big[ b_j (a^{\dagger}_1 +
  (-1)^j e^{i\varphi} a^{\dagger}_2 ) + \mathrm{h.c.} \Big],
\end{equation}
where $b_j$ is the creation operator for photons in mode $j$ of the
fiber, $\nu_j$ is the coupling strength with fiber mode $j$, and the
phase $\varphi$ is induced by the propagation of the field through
the fiber of length $l$: $\varphi = 2\pi \omega l/c$. The
Hamiltonian $H_{IF}$ holds for high finesse cavities and resonant
operations over the time scale much longer than the fiber's
round-trip time \cite{EK+99}. In the short fiber limit
$(2L\bar{\nu})/(2\pi C) \leq 1$, where $L$ is the length of fiber
and $\bar{\nu}$ is the decay rate of the cavity fields into a
continuum of fiber modes, only one resonant mode b of the fiber
interacts with the cavity modes. Therefore, for this case, the
Hamiltonian $H_{IF}$ may be approximated to \cite{SM+06}
\begin{equation}
  \label{eq:Hf}
  H_{IF} = \nu \Big[ b(a^{\dagger}_1 + a^{\dagger}_2 ) +
  \mathrm{h.c.} \Big],
\end{equation}
where the phase $\varphi$ has been absorbed into the annihilation
and creation operators of the mode of the second cavity field.

In the interaction picture, the total Hamiltonian of the
atom-cavity-fiber combined system is
\begin{equation}
  \label{eq:Ht}
  H= \sum_{j=1}^2 \Big( g_j J^-_j a^{\dagger}_j + \mathrm{h.c.}\Big) + \nu
  \Big[ b(a^{\dagger}_1 + a^{\dagger}_2 ) +
  \mathrm{h.c.} \Big].
\end{equation}
We introduce the total excitation operator $\mathcal{N} = N_{1+} +
N_{2+} + a_1^{\dagger}a_1 + a_2^{\dagger}a_2 + b^{\dagger}b$, where
$N_{j+}$ is the number operator of atoms in the excited state in
cavity $j$. It is easily shown that  the excitation operator
commutes with the Hamiltonian \eqref{eq:Ht}. Therefore, the total
excitation number is a conserved quantity \cite{LL+90}.

\section{Quantum state transfer}
We consider that $N$ two-level atoms are trapped in each of the
cavities and take the state $\vert 0, N\rangle$ where $N$ atoms are
all in the ground state as the computational basis $\vert 0\rangle$
and $\vert 1, N-1\rangle$ where $N-1$ atoms are in the ground state
and one atom in the excited state as the computational basis $\vert
1\rangle$. Suppose that at the initial time all modes of both the
cavities and fiber are not excited, and all the atoms in the first
cavity are in a general superposition state of the two basis vectors
$\alpha|0,N\rangle_1 + \beta|1,N-1\rangle_1$, where $\alpha$ and
$\beta$ are complex numbers and are constrained only by the
normalization condition, but all the atoms in the second cavity are
in the state $|0,N\rangle_2$. The goal of quantum state transfer is
to deterministically accomplish the operation:
\begin{equation}
\label{transmission} (\alpha|0,N\rangle_1 +
\beta|1,N-1\rangle_1)\otimes|0,N\rangle_2\rightarrow|0,N\rangle_1
\otimes(\alpha|0,N\rangle_2+\beta|1,N-1\rangle_2).
\end{equation}

The time evolution of the total system is governed by the
Schr\"odinger equation($\hbar=1$)
\begin{equation}
\label{eq:shrodinger}
  i \frac{\partial}{\partial \mathrm{t}} |\Psi (t)\rangle = H
  |\Psi (t)\rangle.
\end{equation}
Suppose that both the cavity and fiber fields are initially in the vacuum
state $|000\rangle_f$. At time t, the
system is in the state $ U(t)|000\rangle_f\otimes(\alpha|0,N\rangle_1 +
\beta|1,N-1\rangle_1)\otimes|0,N\rangle_2=
\alpha|000\rangle_f\otimes|0,N\rangle_1\otimes|0,N\rangle_2 +\beta
U(t)|000\rangle_f\otimes|1,N-1\rangle_1\otimes|0,N\rangle_2,$ where
$U(t)=\exp(-itH)$. The state
$|000\rangle_f\otimes|1,N-1\rangle_1)\otimes|0,N\rangle_2$ belongs to the
subspace with one excitation number, which is spanned by the basis
state vectors
\begin{equation}
\label{eq:swapbasis}
\begin{aligned}
|\phi_1\rangle &= |000\rangle_f |0,N\rangle_1 |1,N-1\rangle_2, \\
|\phi_2\rangle &= |000\rangle_f |1,N-1\rangle_1 |0,N\rangle_2, \\
|\phi_3\rangle &= |001\rangle_f |0,N\rangle_1 |0,N\rangle_2, \\
|\phi_4\rangle &= |010\rangle_f |0,N\rangle_1 |0,N\rangle_2, \\
|\phi_5\rangle &= |100\rangle_f |0,N\rangle_1 |0,N\rangle_2, \\
\end{aligned}
\end{equation}
where $|n_1 n_f n_2\rangle_f$ denotes the field state with $n_1$
photons in the mode of cavity 1, $n_2$ in the mode of cavity 2 and
$n_f$ in the fiber mode, and $|N_{j+},N-N_{j+}\rangle_j$ is the
state of atoms in cavity $j$, in which $N_{j+}$ atoms are in the
excited state and $N-N_{j+}$ atoms in the ground state. A state of
the entire system with one excitation number can be expanded in
terms of the basis vectors (\ref{eq:swapbasis}) as
\begin{equation}
\label{expanssion}
  |\Psi(t)\rangle = \sum_{i=1}^5 C_i (t) |\phi_i\rangle.
\end{equation}
Upon substitution of (\ref{expanssion}) in (\ref{eq:shrodinger}),
Eq. (\ref{eq:shrodinger}) has the matrix form
\begin{equation}
\label{coefficient}
  i \frac{\partial}{\partial \mathrm{t}} C_i (t) = \sum_{j=1}^5 H_{ij}
  C_j (t),
\end{equation}
where $H_{ij}$ are elements of the matrix representation for the
Hamiltonian (\ref{eq:Ht}) in the one-excitation number subspace, i.e.,
\begin{equation}
\label{Hmatrix}
  H =
  \begin{pmatrix}
    0& 0& \sqrt{N}g& 0& 0\\
    0& 0& 0& 0& \sqrt{N}g\\
    \sqrt{N}g& 0& 0& \nu& 0\\
    0& 0& \nu& 0& \nu\\
    0& \sqrt{N}g& 0& \nu& 0
  \end{pmatrix},
\end{equation}
where $g_1=g_2=g$ has been assumed.

The Hamiltonian matrix (\ref{Hmatrix}) has five eigenvalues: ${E}_1
= 0, {E}_{2,3} = \mp \sqrt{N}g, {E}_{4,5} = \mp\sqrt{N g^2 + 2 \nu^2}$. The
corresponding eigenvectors are
\begin{equation}
\label{eigenvectors}
  \begin{aligned}
    |\varphi_1\rangle &= -\frac{r}{\sqrt{1+2 r^2}} |\phi_1\rangle
    -\frac{r}{\sqrt{1+2 r^2}} |\phi_2\rangle + \frac{1}{\sqrt{1+2
        r^2}} |\phi_4\rangle, \\
    |\varphi_2\rangle &= \frac{1}{2} |\phi_1\rangle - \frac{1}{2}
    |\phi_2\rangle - \frac{1}{2} |\phi_3\rangle + \frac{1}{2}
    |\phi_5\rangle, \\
    |\varphi_3 \rangle&= -\frac{1}{2} |\phi_1\rangle + \frac{1}{2}
    |\phi_2\rangle  - \frac{1}{2} |\phi_3\rangle + \frac{1}{2}
    |\phi_5\rangle, \\
    |\varphi_4 \rangle&= -\frac{1}{2\sqrt{1+2 r^2}} |\phi_1\rangle
    -\frac{1}{2\sqrt{1+2 r^2}} |\phi_2\rangle + \frac{1}{2}|\phi_3\rangle
    -\frac{r}{\sqrt{1+2 r^2}} |\phi_4\rangle + \frac{1}{2} |\phi_5\rangle,\\
    |\varphi_5 \rangle&= \frac{1}{2\sqrt{1+2 r^2}} |\phi_1\rangle +
    \frac{1}{2\sqrt{1+2 r^2}} |\phi_2\rangle + \frac{1}{2}|\phi_3\rangle
    +\frac{r}{\sqrt{1+2 r^2}} |\phi_4\rangle + \frac{1}{2}
    |\phi_5\rangle,
  \end{aligned}
\end{equation}
where $r=\nu/(\sqrt{N}g)$.  From (\ref{eigenvectors}), we deduce the
unitary matrix $S$ that diagonalizes the Hamiltonian matrix
(\ref{Hmatrix})
\begin{equation}
  S=
  \begin{pmatrix}
    -\frac{r}{\sqrt{1 + 2r^2}}& -\frac{r}{\sqrt{1 + 2r^2}}& 0&
    \frac{1}{\sqrt{1 + 2r^2}}& 0\\
    \frac{1}{2}& -\frac{1}{2}& -\frac{1}{2}& 0& \frac{1}{2}\\
    -\frac{1}{2}& \frac{1}{2}& -\frac{1}{2}& 0& \frac{1}{2}\\
    -\frac{1}{2\sqrt{1+2r^2}}& -\frac{1}{2\sqrt{1+2r^2}}& \frac{1}{2}&
    -\frac{r}{\sqrt{1+2r^2}}& \frac{1}{2}\\
    \frac{1}{2\sqrt{1+2r^2}}& \frac{1}{2\sqrt{1+2r^2}}& \frac{1}{2}&
    \frac{r}{\sqrt{1+2r^2}}& \frac{1}{2}
  \end{pmatrix}.
\end{equation}
By use of the unitary matrix $S$, Eq. (\ref{coefficient}) can be
rewritten as the compact form
\begin{equation}
\label{D}
  i \frac{\partial}{\partial \mathrm{t}} SC = S H S^{-1} S C,
\end{equation}
where $C=[C_1,C_2,C_3,C_4,C_5]^T$.  Since the matrix $S H S^{-1}$ is
diagonal, a general solution of Eq. (\ref{coefficient}) is given by
\begin{equation}
\label{solution1}
  C_j(t)= \sum_{k=1}^{5} [S^{-1}]_{jk} [SC(0)]_k
  e^{-iE_{k}t}.
\end{equation}
Using this solution, for the initial condition $C(0) =
[0,1,0,0,0]^T$, we have
\begin{equation}
\begin{aligned}
\label{solution2}
  |\Psi(t)\rangle =&U(t)|000\rangle_f \otimes |1,N-1\rangle_1 \otimes
  |0,N\rangle_2\\
  &=\Big[\frac{r^2}{1+2r^2} -\frac{1}{2} \cos(\sqrt{N}gt) +
  \frac{\cos(\sqrt{1+2r^2} \sqrt{N}gt)}{2(1 +2r^2)} \Big]
  |\phi_1\rangle\\
&+ \Big[\frac{r^2}{1+2r^2} +\frac{1}{2} \cos(\sqrt{N}gt) +
\frac{\cos(\sqrt{1+2r^2} \sqrt{N}gt)}{2(1 +2r^2)} \Big]
|\phi_2\rangle\\
 &+ \Big[-\frac{i}{2} \sin(\sqrt{N}gt) +
\frac{i\sin (\sqrt{1+2
    r^2}\sqrt{N} gt)} {2 \sqrt{1+ 2r^2}}\Big] |\phi_3\rangle\\
&+ \Big[- \frac{r}{1+2r^2} + \frac{r \cos (\sqrt{1+2r^2} \sqrt{N}
gt)}{1 +2r^2} \Big]|\phi_4\rangle \\
&+\Big[\frac{i}{2} \sin(\sqrt{N}gt) + \frac{i\sin (\sqrt{1+2
r^2}\sqrt{N}
  gt)} {2 \sqrt{1+ 2r^2}} \Big]|\phi_5\rangle.
\end{aligned}
\end{equation}
From (\ref{solution2}), it is easily shown that at $t =
\pi/(\sqrt{N}g)$ the initial state $|000\rangle_f
\otimes|1,N-1\rangle_1|0,N\rangle_2$ evolves in the state
$|000\rangle_f \otimes|0,N\rangle_1 |1,N-1\rangle_2 $ if the
parameter $r$ fulfills the condition
\begin{equation}
\label{condition}
  r^2 = (4k^2 - 1)/2, ~~k=1,2,3,\cdots.
\end{equation}

Combining the above results together, we have $
U(\pi/\sqrt{N}g)|000\rangle_f\otimes(\alpha|0,N\rangle_1 +
\beta|1,N-1\rangle_1)\otimes|0,N\rangle_2=
\alpha|000\rangle_f\otimes|0,N\rangle_1\otimes|0,N\rangle_2 +\beta
|000\rangle_f\otimes|0,N\rangle_1\otimes|1,N-1\rangle_2.$ Therefore,
the perfect quantum state transfer (\ref{transmission}) is
deterministically implemented. It is noticed that the operation time
of the state transfer is $\pi/(\sqrt{N}g)$. So, the transfer process
can be greatly speeded up with the large number of atoms even if the
coherent interaction strength $g$ is small.

\begin{figure}[htbp]
  \centering
    \includegraphics[width=10cm]{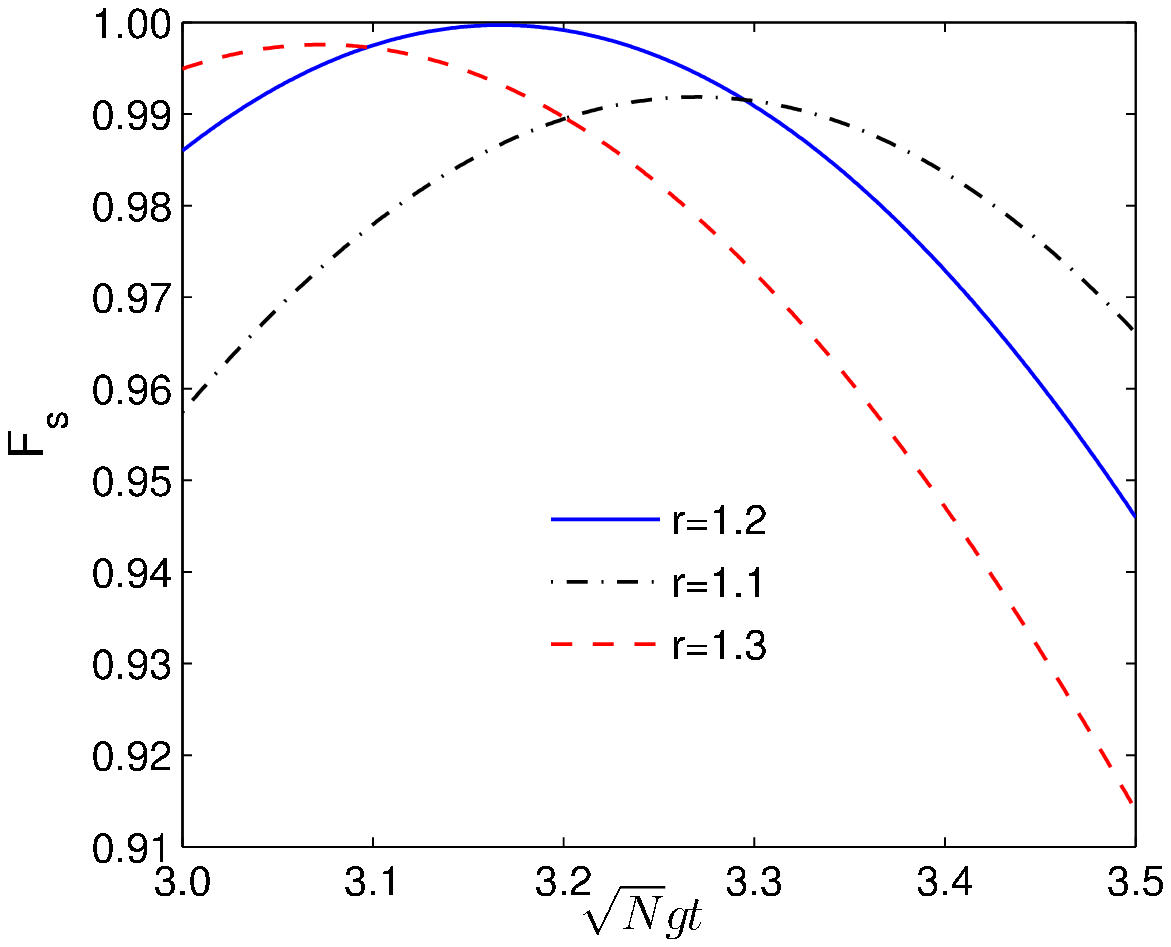}
  \caption{Fidelity of the quantum state transfer as a function of $\sqrt{N}gt$.}
\label{fig:swap}
\end{figure}

If the condition (\ref{condition}) is strictly satisfied by the
coupling constants, the ideal quantum state transfer can be
implemented. In practical situations, however, the mismatch between
the coupling strengths from the condition (\ref{condition})
inevitably happens. From \eqref{transmission}, the target state of
the transmission is $|\psi_s\rangle =
|000\rangle_f\otimes|0,N\rangle_1 \otimes (\alpha |0,N\rangle_2 +
\beta |1,N-1\rangle_2)$. Fidelity of the quantum state transfer is
defined as $F_s = |\langle \Psi(t) | \psi_s \rangle|^2$. In Fig.
\ref{fig:swap}, the fidelity is shown for the state transfer
$(|0,N\rangle_1+|1,N-1\rangle_1)/\sqrt{2}\otimes|0,N\rangle_2\rightarrow|0,N\rangle_1
\otimes(|0,N\rangle_2+|1,N-1\rangle_2)/\sqrt{2}$ with different
values of the parameter $r$ around $\sqrt{1.5}$ with which the
perfect state transfer is implemented. One can observe that the
fidelity is highly stable to the mismatch of the coupling strengths
from the condition (\ref{condition}). In Fig. \ref{fig:swap}, one
may also observe that the fidelity varies smoothly as a function of
the dimensionless time. This feature is useful for switching off the
interaction between the atoms and the fields with control pulses
once the quantum state transfer is achieved.

\section{Quantum swap gate}

Suppose that at the initial time  all the atoms in cavity $j$ are in
a general superposition state: $\alpha_j |0,N\rangle_j + \beta_j
|1,N-1\rangle_j$ and  all modes of both the cavities and fiber are
in the vacuum state. Our goal is to deterministically realize the
state swap between the two atomic systems: $(\alpha_1|0,N\rangle_1 +
\beta_1 |1,N-1 \rangle_1) \otimes (\alpha_2|0,N\rangle_2 +
\beta_2|1,N-1\rangle_2) \rightarrow (\alpha_2|0,N\rangle_1 +
\beta_2|1,N-1\rangle_1) \otimes (\alpha_1|0,N\rangle_2 + \beta_1
|1,N-1\rangle_2)$.

In the state swap, the three types of atomic states are involved:
$(1) |0,N\rangle_1|0,N\rangle_2$; $(2)|1,N-1 \rangle_1
|0,N\rangle_2$, and $|0,N \rangle_1 |1,N-1 \rangle_2$;
$(3)|1,N-1\rangle_1 |1,N-1 \rangle_2$, which belong to subspaces
with zero-, one- and two- excitation numbers, respectively. Since
the Hamiltonian (\ref{eq:Ht}) conserves the total excitation number,
Eq. (\ref{eq:shrodinger}) can be solved in each of the subspaces,
respectively. When non photons are initially in both the cavities
and fiber, the state $|0,N\rangle_1
|0,N\rangle_2\otimes|000\rangle_f$ is unchanged. For the initial
state $(\beta_1\alpha_2|1,N-1 \rangle_1
|0,N\rangle_2+\beta_2\alpha_1|0,N \rangle_1 |1,N-1
\rangle_2)\otimes|000\rangle_f$, Eq. (\ref{eq:shrodinger}) is solved
in the subspace spanned by the basis vectors (\ref{eq:swapbasis}).
The two-excitation subspace is spanned by the basis vectors
\begin{equation}
\label{eq:swapbasis1} |\phi_{n_1n_fn_2}^{m_1m_2} \rangle = |m_1,
N-m_1\rangle_1 |m_2, N-m_2\rangle_2\otimes|n_1n_f n_2\rangle_f,
\end{equation}
with the conditions $0 \leq n_1,n_f,n_2,m_1,m_2 \leq 2$ and
$n_1+n_f+ n_2 + m_1 +m_2 = 2$. In this subspace, the Dicke states
$|\varphi_{000}^{20} \rangle$ and $|\varphi_{000}^{02} \rangle$ in
which two atoms in the same cavity are simultaneously excited are
involved.
\begin{figure}[htbp]
  \centering
    \includegraphics[width=10cm]{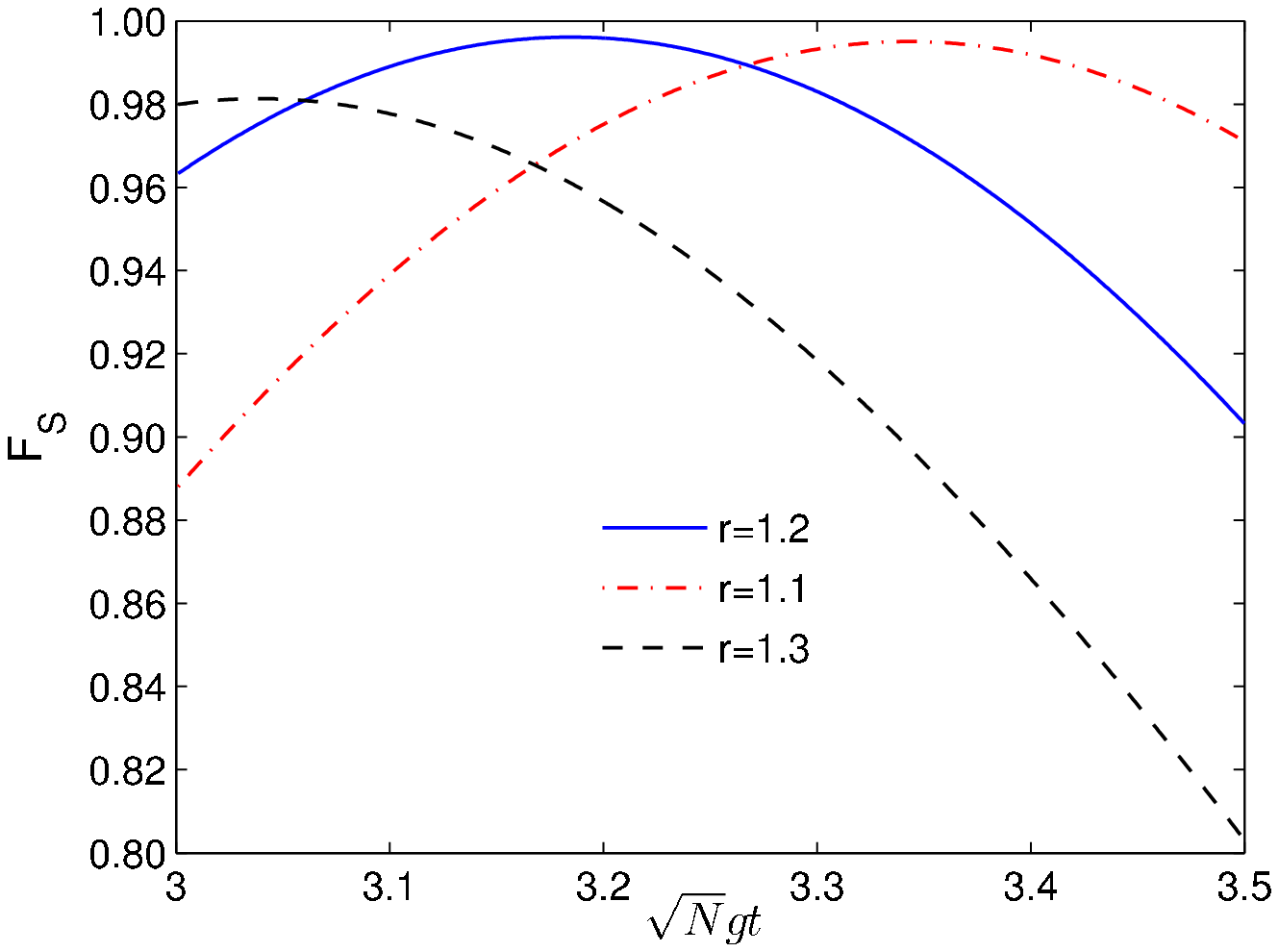}
  \caption{The average fidelity of the swap gate as a function of  $\sqrt{N}gt$.}
\label{fig:Gswap}
\end{figure}
Here, we assume that the ``dipole blockade'' effect takes place,
which was proposed by \citet{LF+01} and observed in the recent
experiments\cite{TF+04,SR+04,LR+05,VV+06}. This effect can highly depress
the transition from single- to double excitation Dicke states. For
the initial state $|1,N-1 \rangle_1|1,N-1
\rangle_2\otimes|000\rangle_f$, we solve Eq. (\ref{eq:shrodinger})
in the subspace spanned by the basis vectors (\ref{eq:swapbasis1})
with neglecting the double excitation Dicke states. On combining the
results obtained from the zero-, single- and two-excitation
subspaces, we can find the state $| \Psi(t) \rangle$ of the entire
system at time t.

If $\alpha_i= \sin \theta_i$ and $\beta_i = \cos \theta_i$, the target
state $|\Psi_s\rangle$ is $
(\sin\theta_2|0,N\rangle_1 + \cos\theta_2|1,N-1\rangle_1) \otimes
(\sin\theta_1|0,N\rangle_2 +
\cos\theta_1|1,N-1\rangle_2)\otimes|000\rangle_f$. The average fidelity of
the swap gate is defined as
\begin{equation}
F_S =\frac{1}{4\pi^2}\int_0^{2\pi}d\theta_1\int_0^{2\pi}d\theta_2
\vert\langle \Psi_s |\Psi(t) \rangle\vert^2.
\end{equation}
In Fig. \ref{fig:Gswap}, the average fidelity is plotted as a
function of $\sqrt{N}gt$ for various values of the parameter $r$. It
is observed that a highly reliable swap gate with the fidelity
larger than $0.995$ can be deterministically achieved at
$\sqrt{N}gt\simeq 3.2$ around $r\simeq 1.2$. Since the operation
time of the swap gate is proportional to $1/\sqrt{N}$,  the swap
gate can be greatly speeded up when the number of atoms is large. In
Fig. \ref{fig:Gswap}, one may also observe that the maximum of the
fidelity is relatively stable with respect to the dimensionless time
$\sqrt{N}gt$ and the variation of the coupling constants.

\section{Quantum Entangling Gate}

In this section, we investigate to create the entangled states :
$(|1,N-1\rangle_1 |0,N\rangle_2\pm|0,N\rangle_1
|1,N-1\rangle_2)/\sqrt{2}$.

In the Schr\"odinger picture, the Hamiltonian \eqref{eq:Ht} has the
form
\begin{equation}
 \label{eq:Hent}
  H = \omega (a^{\dagger}_1 a_1 + a^{\dagger}_2 a_2 + b^{\dagger}
  b + J^z_1 + J^z_2)+H_{AF}+H_f,
\end{equation}
where $\omega$ is frequency of the transition between the excited
and ground states of the two-level atom, and $J^z_j= \sum_{i=1}^N
\sigma_{i}^z(j)$ with $\sigma_{i}^z(j) = (|e_{i}\rangle \langle
e_{i}| - |g_{i}\rangle \langle g_{i}|)/2$ for atom i in cavity $j$.
Here, it is assumed that the interaction between the atoms and the
cavity field, and the coupling between the cavity and fiber fields
are on resonance. By use of the canonical transformations
\cite{SM+06}
\begin{equation}
\label{eq:a2c}
 \begin{aligned}
  a_1&= \frac{1}{2} (c_+ + c_- + \sqrt{2} c),\\
  a_2&= \frac{1}{2} (c_+ + c_- - \sqrt{2} c), \\
  b&= \frac{1}{\sqrt{2}} (c_+ - c_-),
 \end{aligned}
\end{equation}
three normal bosonic modes $c$ and $c_{\mp}$ are introduced. In
terms of these new bosonic operators, the Hamiltonian
(\ref{eq:Hent}) can be expressed as
\begin{equation}
\label{hc}
  \begin{aligned}
  H = &\frac{1}{2}\Big[  g_1 J^+_1 ( c_+ + c_- + \sqrt{2} c) +
  g_2 J_2^{+} ( c_+ + c_- - \sqrt{2} c) + \mathrm{h.c.}
  \Big] \\ &+\omega (c^{\dagger}c + J^z_1+J^z_2) + (\omega + \sqrt{2}
  \nu) c_+ c_+^{\dagger} + (\omega - \sqrt{2}\nu) c_- c_-^{\dagger}.
  \end{aligned}
\end{equation}
It is seen that frequencies of the normal mode $c$ and $c_{\mp}$ are
$\omega$ and $\omega \mp \sqrt{2}\nu$, respectively. Therefore, the
mode c resonantly interacts with the atoms but the modes $c_{\mp}$
non-resonantly interact with the atoms. For $\nu \gg \sqrt{N}|g_j|$,
excitations of the nonresonant modes can be highly suppressed. In
this case, the modes $c_{\mp}$ can be safely neglected. In this way,
the system reduces to two qubits resonantly coupled through a
single-mode of the cavity field, and the Hamiltonian (\ref{hc}) in
the interaction picture becomes
\begin{equation}
\label{eq:Hint}
  H=\frac{1}{\sqrt{2}} \Big( g_1 J_1^- c^{\dagger} - g_2 J_2^- c^{\dagger} +
  \mathrm{h.c.} \Big).
\end{equation}

Suppose that the atoms are in the state
$|1,N-1\rangle_1|0,N\rangle_2$, and the mode $c$ is in the vacuum
state at the initial time. From (\ref{eq:Hint}), the system at the
later time is restricted to the subspace spanned by the basis
vectors
\begin{equation}
\label{basis1}
  \begin{aligned}
    |\phi_1\rangle &= |0\rangle_c |0,N\rangle_1 |1,N-1\rangle_2,\\
    |\phi_2\rangle &= |0\rangle_c |1,N-1\rangle_1 |0,N\rangle_2,\\
    |\phi_3\rangle &= |1\rangle_c |0,N\rangle_1 |0,N\rangle_2,
  \end{aligned}
\end{equation}
where $|0\rangle_c$ and $|1\rangle_c$ are number states of the
normal modes $c$ with zero and one photon, respectively. Using the
same method as in section III, one can show that at time $t$ the
system evolves in the state
\begin{equation}
  \begin{aligned}
\label{eq:Psi}
    |\Psi(t)\rangle &= \frac{g_1^2 +
      \cos(\sqrt{N(g_1^2 + g_2^2)/2}~t)g_2^2}{g_1^2 +g_2^2}
    |\phi_1\rangle + \frac{g_1g_2 - \cos(\sqrt{N(g_1^2 +
        g_2^2)/2}~t)g_1g_2}{g_1^2 +g_2^2} |\phi_2\rangle \\ &+
    i\frac{g_1\sin(\sqrt{N(g_1^2 + g_2^2)/2}~t)}{\sqrt{g_1^2 + g_2^2}}
    |\phi_3\rangle.
  \end{aligned}
\end{equation}
If requiring that the amplitude of $|\phi_3\rangle$ vanishes and the
absolute values of the amplitudes of $|\phi_{1,2}\rangle$ in
(\ref{eq:Psi}) are equal, we obtain the results:
 \begin{list}{}{}
\item \textbf{(a)}  if $g_2=(1 + \sqrt{2}) g_1$ and $\sqrt{N}g_1t =
\pi/\sqrt{2+\sqrt{2}}$, the field comes back to the vacuum state and
the atoms are in the entangled state $|\psi_{E1} \rangle =
(|1,N-1\rangle_1 |0,N\rangle_2 + |0,N\rangle_1
|1,N-1\rangle_2)/\sqrt{2}$;
\item \textbf{(b)} if $g_2=(-1 + \sqrt{2}) g_1$ and $\sqrt{N}g_1t =
\pi/\sqrt{2-\sqrt{2}}$, the field comes back to the vacuum state and
the atoms are in the entangled state $|\psi_{E2} \rangle =
(|1,N-1\rangle_1 |0,N\rangle_2 - |0,N\rangle_1
|1,N-1\rangle_2)/\sqrt{2}$.
\end{list}
If $N_1 \neq N_2$, we also find that the entangled states can be
generated if either $\sqrt{N_2}g_2 = (1+\sqrt{2}) \sqrt{N_1}g_1$ or
$\sqrt{N_2}g_2=(-1+\sqrt{2}) \sqrt{N_1} g_1$ is fulfilled.

\begin{figure}[htbp]
  \centering
\includegraphics[width=8cm]{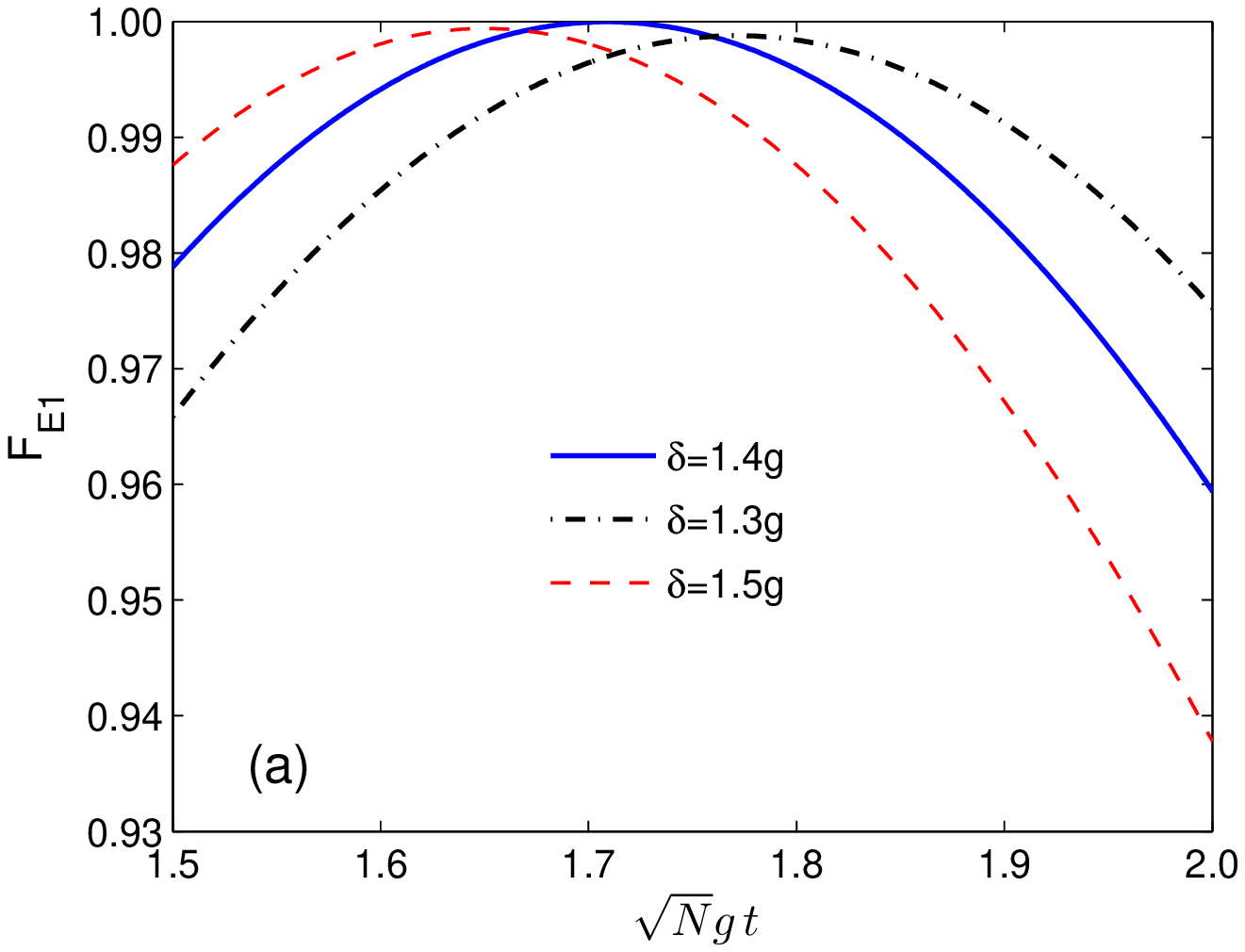}
\includegraphics[width=8cm]{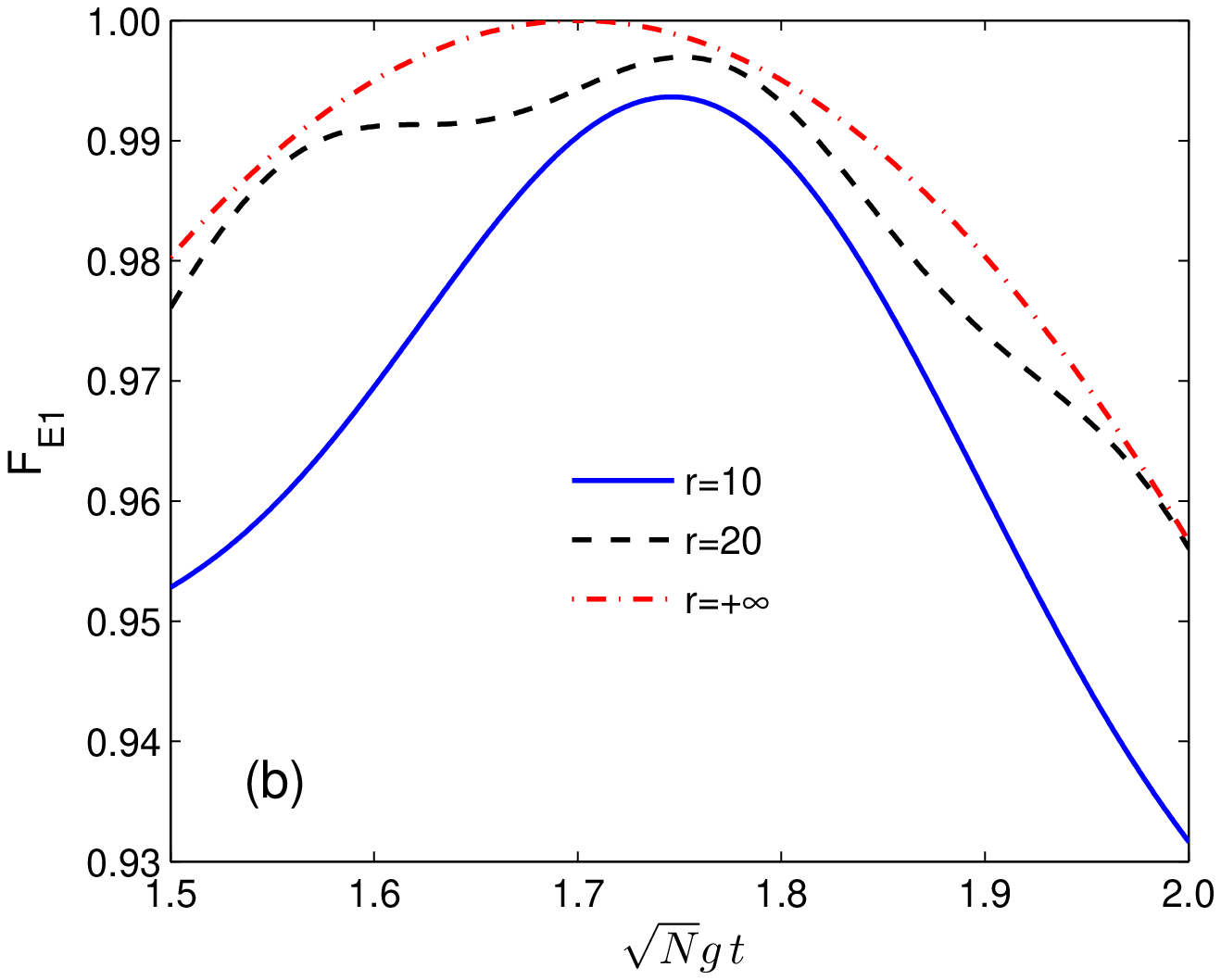}
  \caption{Fidelity of the entangling gate versus
  $\sqrt{N}gt$. (a) $r=\infty$;(b)$\delta=\sqrt{2}g$.}
\label{fig:entangle}
\end{figure}

In practical situations, the ratio of the coupling constants
required for realizing the entangling gate may not be satisfied
exactly. In order to see the effects of the imperfections in the
coupling constants, we define  fidelity of the entangling gate:
$F_{Ei} = |\langle \Psi(t) | \psi_{Ei} \rangle|^2$. In Fig.
\ref{fig:entangle}(a), setting $g_1=g$ and $\delta = g_2 - g_1$, we
plot the fidelity as a function of the dimensionless time
$\sqrt{N}gt$ for different values of the parameter $\delta$. It is
seen that the fidelity is remarkably robust to the deviation of the
coupling constants from the values for the perfect entangling gate.

In order to check in which regime the non resonant normal modes can
be safely neglected, we directly solve Eq. (\ref{eq:shrodinger}) in
the subspace spanned by the basis vectors \eqref{eq:swapbasis}. In
the calculation, we suppose that the atoms are initially in
$|1,N-1\rangle_1 |0,N \rangle_2$ and all the field modes are in the
vacuum state. In Fig. \ref{fig:entangle}(b), the fidelity is plotted
for different values of the parameter $r$. It is observed that the
maximum of the fidelity can be larger than $0.99$ if $r$ is beyond
$20$.

Similar to the swap gate, the operation time of the entangling gate
is also proportional to $1/\sqrt{N}$. Therefore, the entangling gate
can be greatly speeded up as the number of atoms increases.

\section{Controlled-Z gate}
\label{sec:CNOT}

In this section, we investigate how to realize the controlled-Z gate
between the two atomic systems: $(|0,N\rangle_1 + |1,N-1\rangle_1)
\otimes (|0,N\rangle_2 + |1,N-1\rangle_2)/2\rightarrow(|0,N\rangle_1
|0,N\rangle_2 + |0,N\rangle_1 |1,N-1\rangle_2 + |1,N-1\rangle_1
|0,N\rangle_2 - |1,N-1\rangle_1 |1,N-1\rangle_2)/2$ \cite{NC00}. A
controlled-Z gate is important in building quantum computers since a
CNOT gate \cite{BD+95} can be constructed from one controlled-Z gate
and two Hardamard gates \cite{NC00}.

As in the proceeding section, we assume that the limit $\nu \gg
\sqrt{N}|g_j|$ is valid. In the controlled-Z gate operation, when
the resonant c-mode is initially in the vacuum state, the three
types of atom states are involved: $(1) |0,N\rangle_1|0,N\rangle_2$;
$(2) |1,N-1\rangle_1|0,N\rangle_2$ and
$|0,N\rangle_1|1,N-1\rangle_2$; $(3)
|1,N-1\rangle_1|1,N-1\rangle_2$, which belong to subspaces with the
zero-, single- and two- excitation numbers, respectively. The zero
subspace contains only one state: $|0,N\rangle_1|0,N\rangle_2$. The
single-excitation subspace is spanned by the basis vectors
(\ref{basis1}). The two-excitation subspace is spanned by the basis
vectors
\begin{equation}
\label{eq:CNOT2}
 \begin{aligned}
  |\phi_1\rangle &= |0\rangle_c |1,N_1-1\rangle_1|1,N_2-1\rangle_2,\\
  |\phi_2\rangle &= |1\rangle_c |0,N_1\rangle_1 |1,N_2-1\rangle_2, \\
  |\phi_3\rangle &= |1\rangle_c |1,N_1-1\rangle_1 |0,N_2\rangle_2, \\
  |\phi_4\rangle &= |2\rangle_c |0,N_1\rangle_1 |0,N_2\rangle_2,\\
  |\phi_5\rangle &= |0\rangle_c |2,N_1-2\rangle_1 |0,N_2\rangle_2,\\
  |\phi_6\rangle &= |0\rangle_c |0,N_1\rangle_1 |2,N_2-2\rangle_2.
 \end{aligned}
\end{equation}
In the last two vectors, two atoms in the same cavity are excited.
Here, we assume that the "dipole blockade" effects \cite{LF+01} take
place. Then, the double-excitation Dicke states can be neglected.

Since the total excitation number $c^{+}c+N_1+N_2$ is conserved in
the coherent evolution governed by the Hamiltonian (\ref{eq:Hint}),
using the same method in section III,  we solve the Schr\"{o}dinger
equation with the Hamiltonian (\ref{eq:Hint}) in the zero-, single-
and two-excitation subspaces for the initial states
$|0,N\rangle_1|0,N\rangle_2, (|1,N-1\rangle_1|0,N\rangle_2+
|0,N\rangle_1|1,N-1\rangle_2)/\sqrt{2}$, and
$|1,N-1\rangle_1|1,N-1\rangle_2$, respectively. On combining the
results obtained from the subspaces, we can find the state
$|\Psi(t)\rangle$ of the entire system at time $t$ with the initial
condition $|\Psi(0)\rangle=|0\rangle_c\otimes(|0,N\rangle_1 +
|1,N-1\rangle_1) \otimes (|0,N\rangle_2 + |1,N-1\rangle_2)/2$. For
the controlled-Z gate, the target state is $|\Psi_Z\rangle =
|0\rangle_c\otimes(|0,N\rangle_1 |0,N\rangle_2 + |0,N\rangle_1
|1,N-1\rangle_2 + |1,N-1\rangle_1 |0,N\rangle_2 - |1,N-1\rangle_1
|1,N-1\rangle_2)/2$. Then, the fidelity of the controlled-Z gate is
defined as $F_{Z} =| \langle \Psi_Z |\Psi(t) \rangle|^2$.

\begin{figure}[htbp]
  \centering
   \includegraphics[width=8cm]{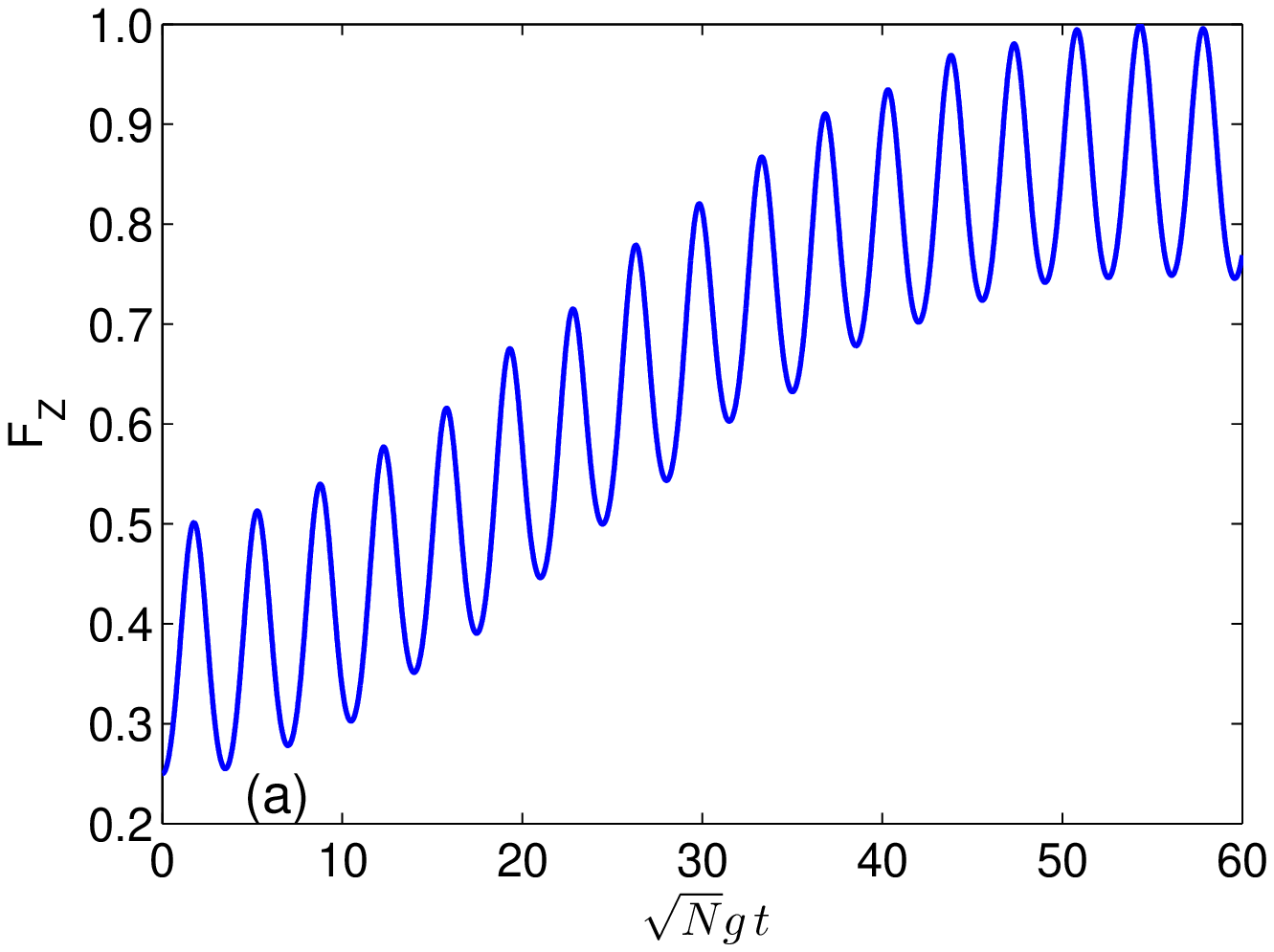}
   \includegraphics[width=8cm]{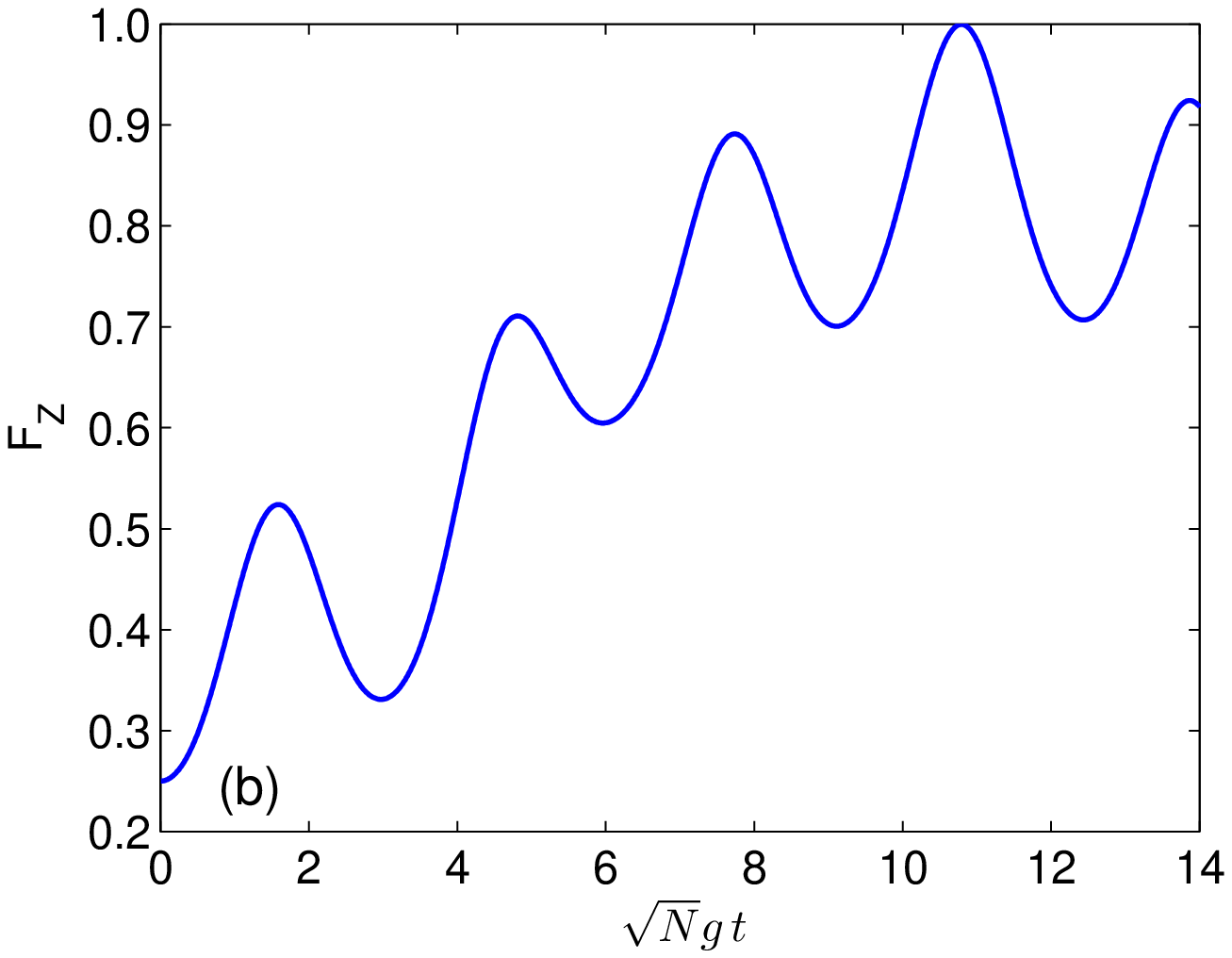}
   \includegraphics[width=8cm]{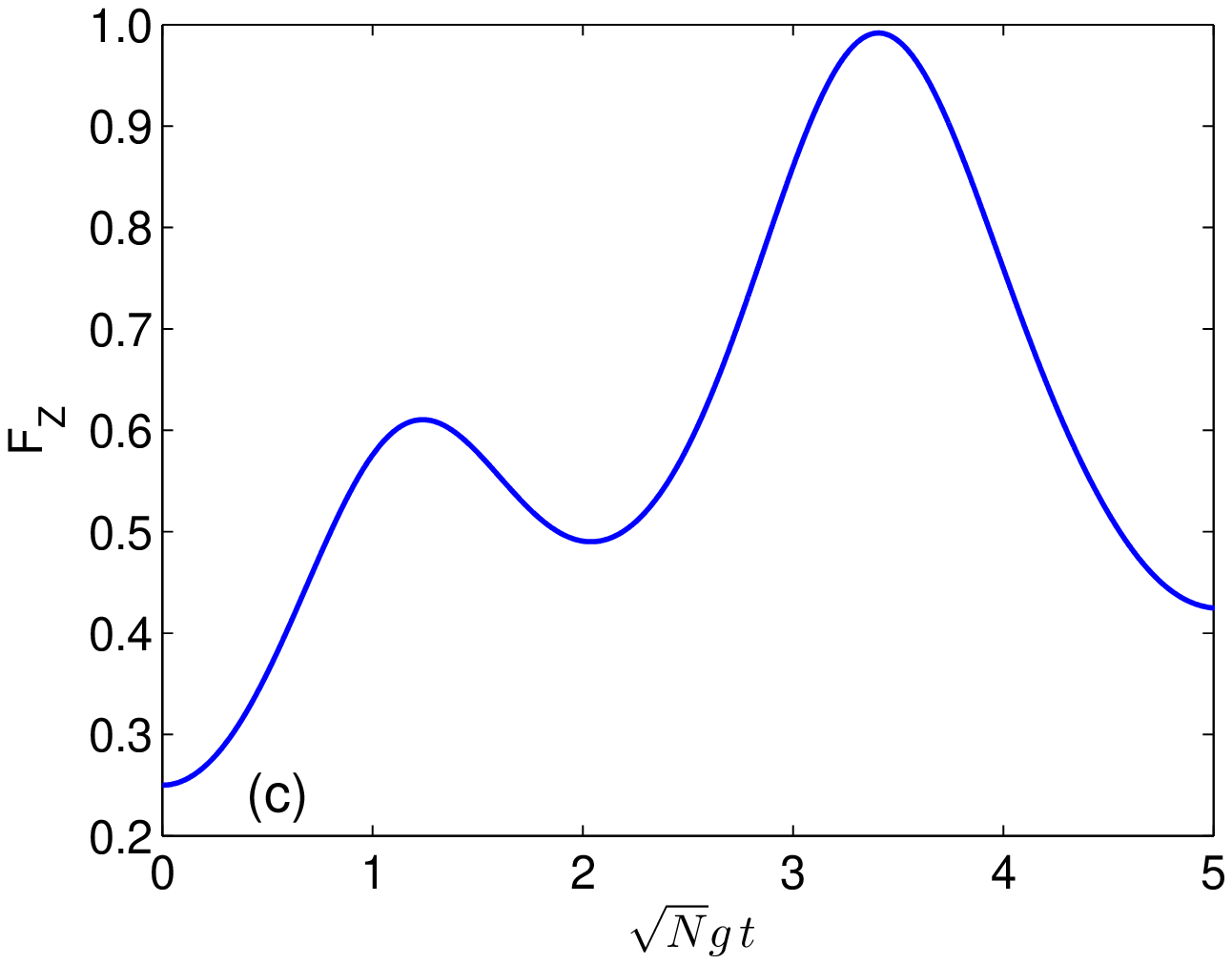}
  \caption{Fidelity of the controlled-Z gate as a function of $\sqrt{N}gt$
   with $\delta=$ (a) $0.07g$; (b) $0.35g$; (c) $1.2g$.} \label{fig:CNOT}
\end{figure}

\begin{figure}[htbp]
  \centering
\includegraphics[width=8cm]{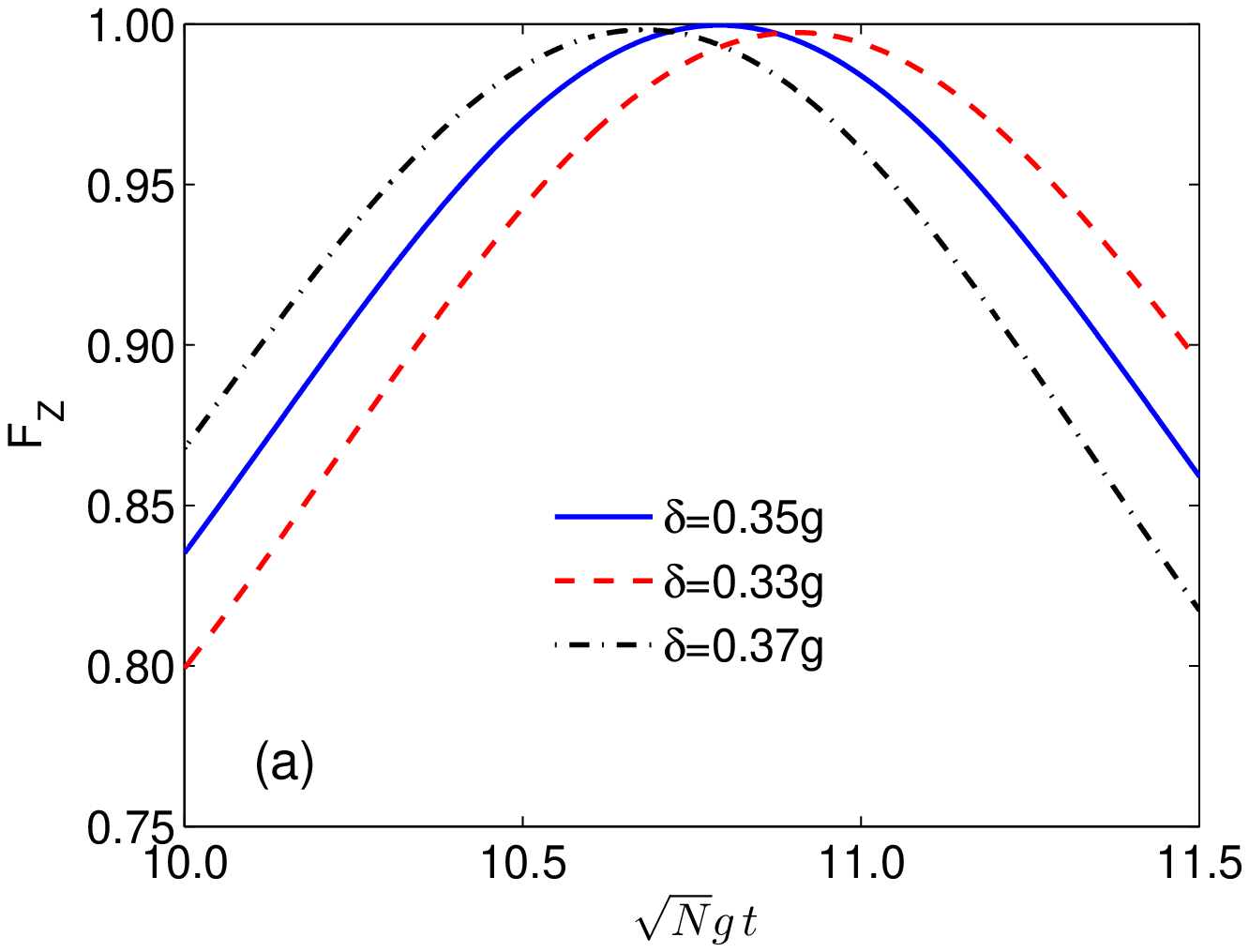}
\includegraphics[width=8cm]{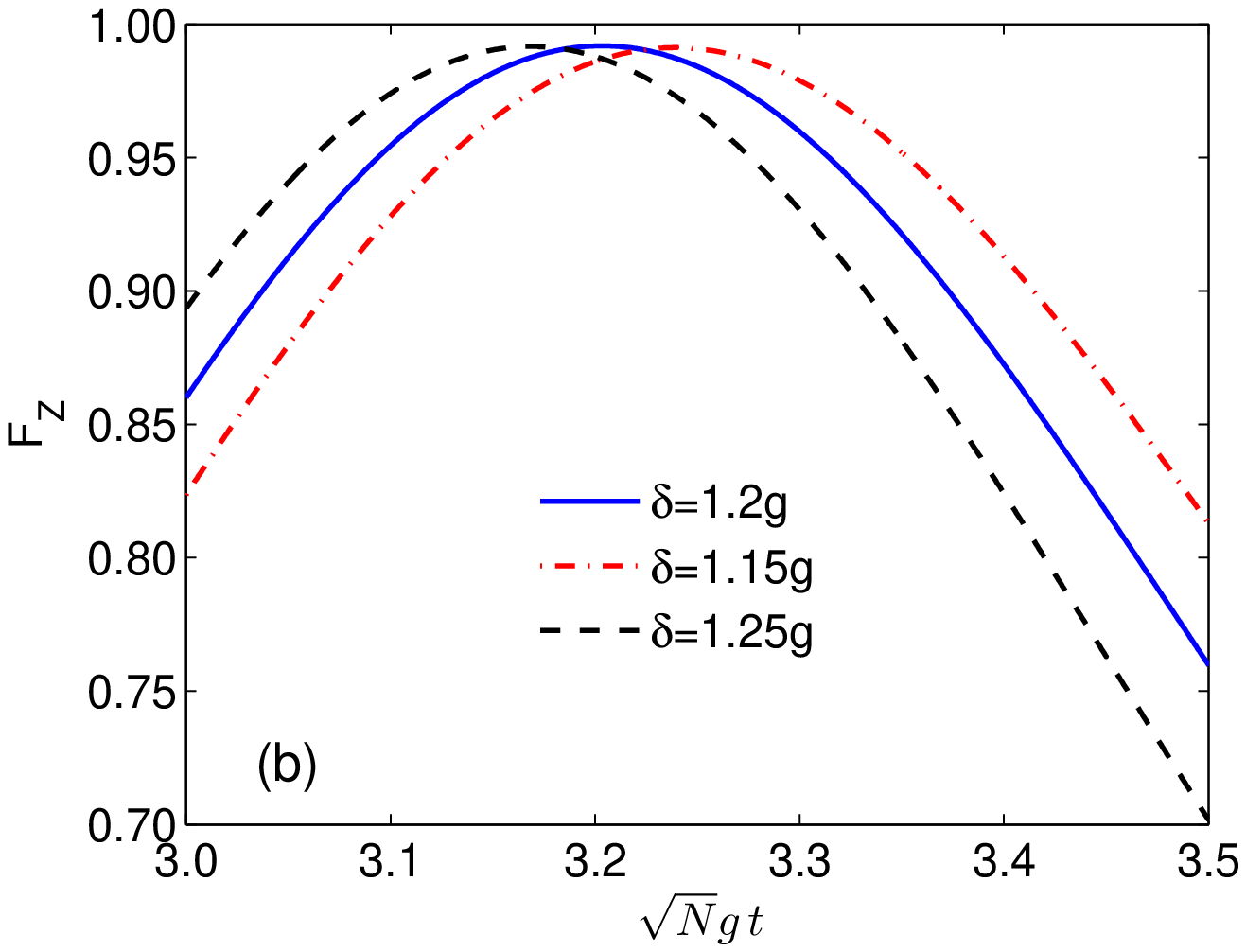}
\includegraphics[width=8cm]{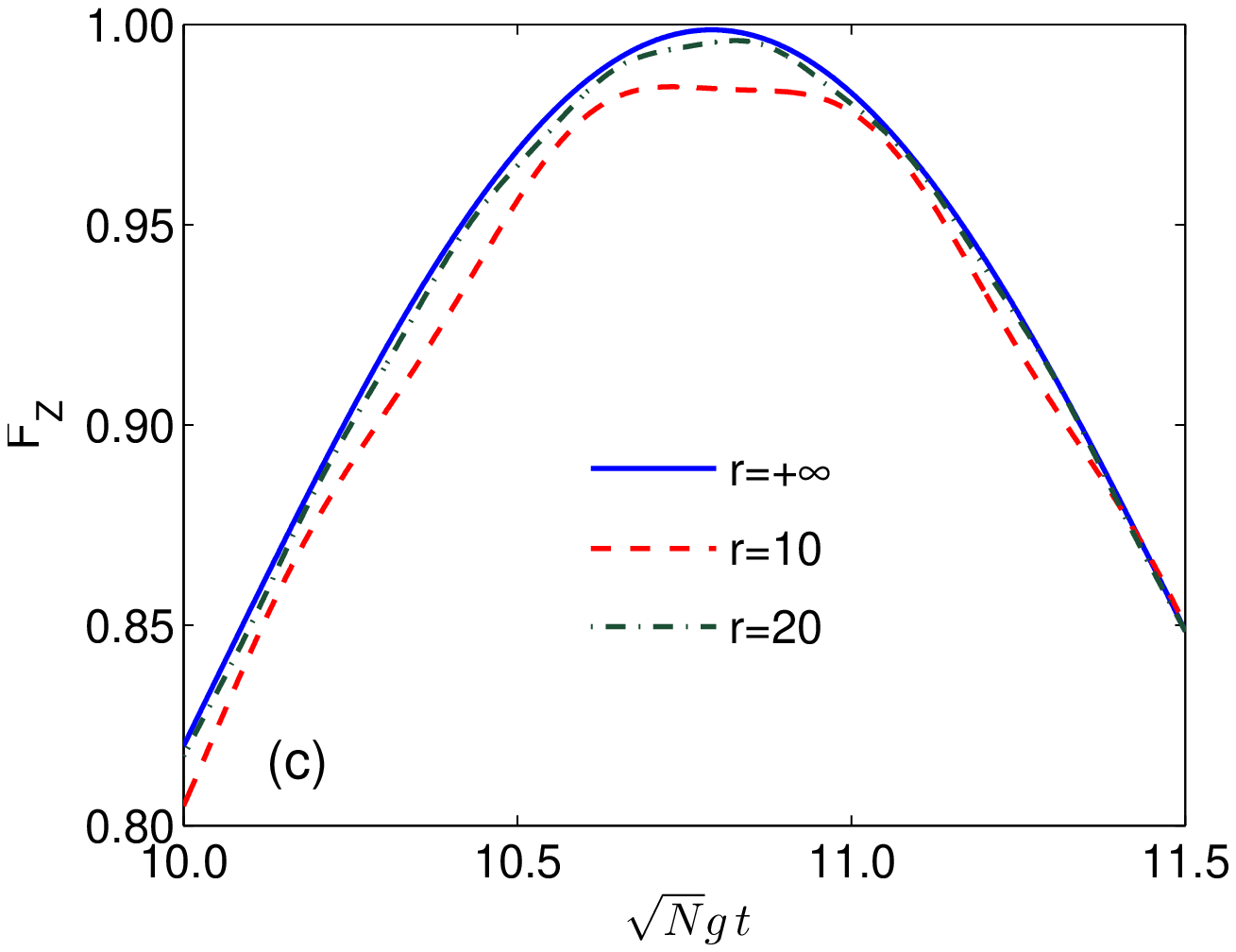}
   \caption{Fidelity of the controlled-Z gate as a
     function of the dimensionless time $\sqrt{N}gt$.}
\label{fig:CNOT1}
\end{figure}

In the calculation, as section in V, let's set $g_1=g$ and introduce
the parameter $\delta=g_2-g_1$. The controlled-Z gate can be
realized by properly choosing the parameter $\delta$. As shown in
Fig. \ref{fig:CNOT}, for $\delta\simeq 0.07g$, the controlled-Z gate
with the fidelity larger than $0.999$ is achieved at $\sqrt{N}gt
\simeq 54.3 $.  If $\delta=0.35g$, the controlled-Z gate with the
fidelity bigger than $0.999$ reaches at $\sqrt{N}gt\simeq 10.80$. If
$\delta=1.20$, the controlled-Z gate with the fidelity larger than
$0.99$ is realized at $\sqrt{N}gt \simeq 3.4$ . In the calculation,
it is noticed that the controlled-Z gate is speeded up as the
parameter $\delta$ increases. The sensitivity of the fidelity to the
difference between the atom-field couplings has also been checked.
With comparing Fig. \ref{fig:CNOT1}(a) to Fig. \ref{fig:CNOT1}(b),
one may find that the larger $\delta$ is and the more insensitive
the fidelity to the coupling difference is. In the non-resonant and
one-atom case \cite{SM+06}, it was also pointed out that the larger
$\delta$ is and the less the operation time of the controlled-Z gate
is. However, the fidelity and the stability of the quantum gate
greatly decrease as the parameter $\delta$ increases. In the
resonant and multiatom case under consideration, as the parameter
$\delta$ increases, the stability of the quantum gate is highly
enhanced but meanwhile the fidelity decreases only from $0.999$ to
$0.99$. In Fig. \ref{fig:CNOT1}(c), the results obtained by directly
solving the Schr\"{o}dinger equation \eqref{eq:shrodinger} with
$\delta=0.35g$ are shown. It is noticed that the off-resonant modes
can be safely neglected in implementing the controlled-Z gate when
$r=\nu/(\sqrt{N}g)>20$.

In the present scheme, all the atoms interact resonantly with the
cavity fields. We find that the resonant interaction can greatly
speed up the operation of the controlled-Z gate. For example, if
$\delta=1.2g$ and $N=1$, the operation time needed to implement the
controlled-Z gate is only $1/8$ of the time needed in the
nonresonant case that was considered in\cite{SM+06}. Meanwhile, in
the multiatom case, the operation time of the controlled-Z gate is
proportional to $1/\sqrt{N}$. Therefore, we expect that the
controlled-Z gate based on the multi atoms and the resonant
interaction can run much faster than on the single atom and the
nonresonant interaction. In this way, the influence of dissipations
such as spontaneous emission and cavity and fiber losses on the
operation of the controlled-Z gate may be greatly diminished.

\section{Effects of spontaneous emission and photon leakage}

In this section, we investigate the influence of atomic spontaneous
emission and photon leakage out of the cavities and fiber on the
quantum state transfer, and the swap, entangling and controlled-Z
gates which have been discussed in the previous sections.

In the present calculation, we assume that the atoms interact
collectively with the privileged local cavity modes but individually
with other modes of the electromagnetic field. This scheme may be
realized by properly arranging the spatial distribution of atoms in
cavity. For example, the spatial distribution shape of atoms may be
made so narrow along the axial direction of the cavity that all the
atoms see the same field and collectively interact with the cavity
field but so wide along the transversal direction that all the atoms
can individually interact with all modes of the electromagnetic
field except the cavity one. We will later discuss this point in
detail and show this arrangement can be realized with present cavity
and trap techniques. Based on the consideration, the master equation
of motion for the density matrix of the entire system may be written
as
\begin{equation}
\label{eq:master}
\begin{aligned}
  \dot{\rho} = &-i[H,\rho] + \gamma \sum_{j=1}^2 L[a_j]\rho + \beta
  L[b]\rho\\
  &+\kappa\sum_{j=1}^2 \sum_{i=1}^N \Big [2 \sigma_{i}^-(j) \rho
\sigma_{i}^+(j) - \sigma_{i}^+(j) \sigma_{i}^-(j) \rho - \rho
\sigma_{i}^-(j) \sigma_{i}^+(j)\Big],
\end{aligned}
\end{equation}
where $L[o]\rho=2o\rho o^{\dagger}-o^{\dagger}o\rho -\rho
o^{\dagger}o$, and $\kappa$, $\gamma$ and $\beta$ are rates,
respectively, for spontaneous emission of the atoms, photon leakage
out of the cavity and fiber. For simplicity, we assume the rates
equal for both the cavities and the mean number of thermal photons
in both the cavities and fiber is zero.

\begin{figure}[htbp]
  \centering
\includegraphics[width=8cm]{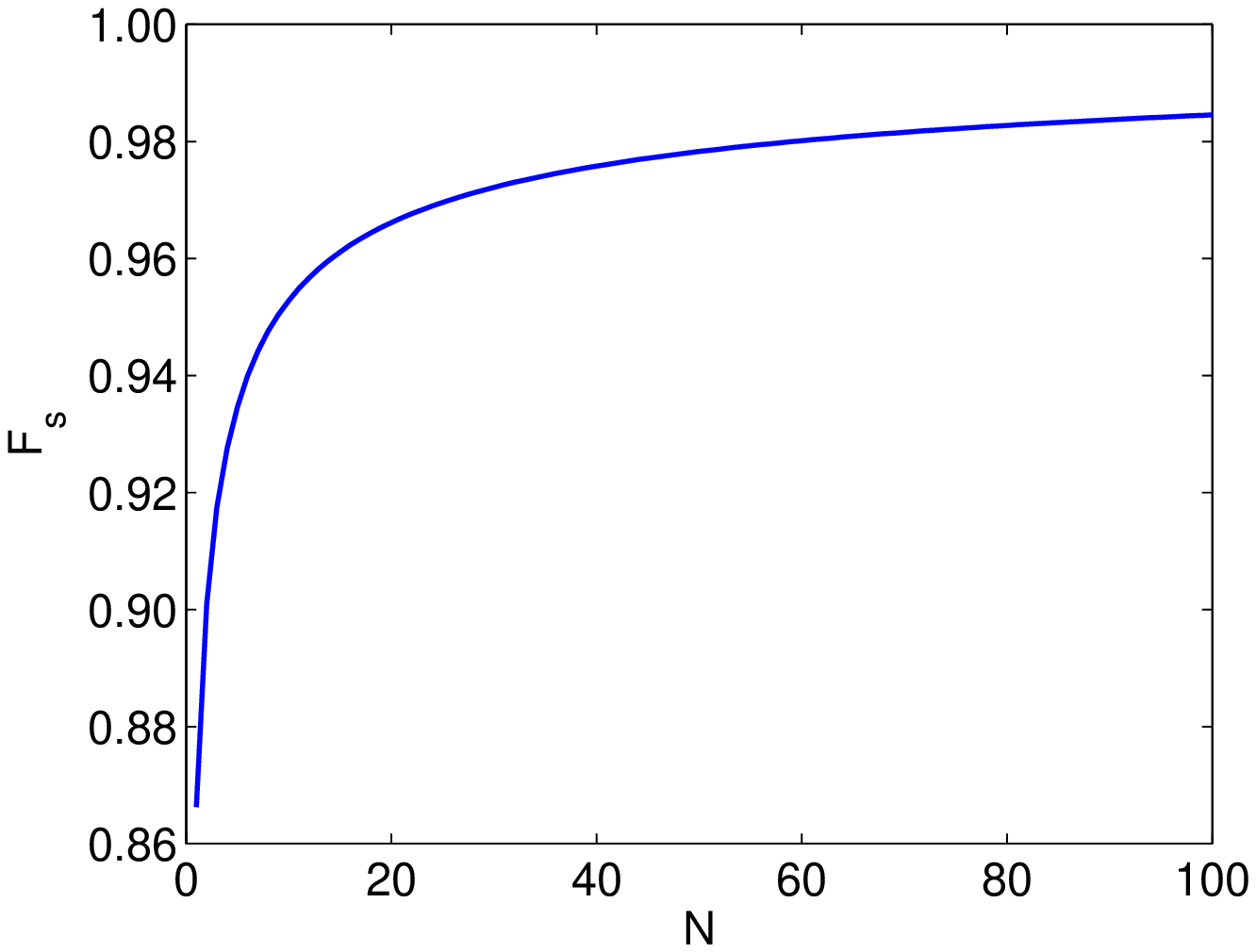}
\includegraphics[width=8cm]{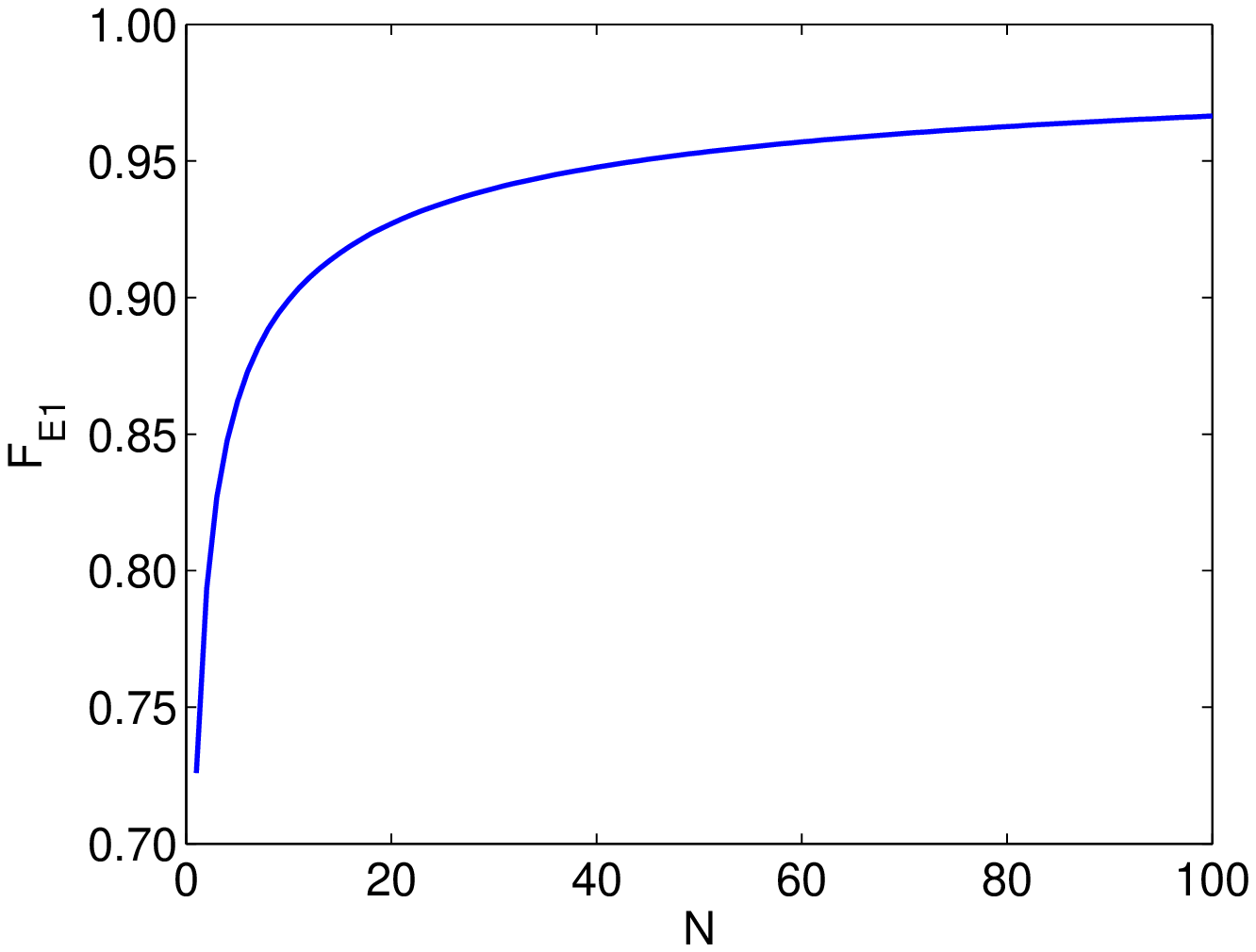}
  \caption{Fidelity of (a) the state transfer and (b) the entangling
    gate as a function of the number of atoms.}
  \label{fig:diss}
\end{figure}

At first, let's consider the state transfer: $(|0,N\rangle_1 +
|1,N-1\rangle_1)/\sqrt{2}\otimes|0,N\rangle_2\rightarrow|0,N\rangle_1
\otimes(|0,N\rangle_2+|1,N-1\rangle_2)/\sqrt{2}$. In this case,
 the master equation (\ref{eq:master}) is numerically solved in the
space spanned by the basis vectors (\ref{eq:swapbasis}) and
$|000\rangle_f|0,N\rangle_1|0,N\rangle_2$. In the calculation, $g_1=g_2=g$ and
$\kappa=\gamma=\beta=0.1g$ are chosen. For the state transmission
under consideration, the target state is
$|\Psi_s\rangle=|000\rangle_f\otimes|0,N\rangle_1\otimes
(|0,N\rangle_2+|1,N-1\rangle_2)/\sqrt{2}$. Fidelity of the state
transfer is defined as $F_s(t)=\langle\Psi_s\vert\rho(t)|\Psi_s\rangle$. In Fig.
\ref{fig:diss}(a), the maximum of the fidelity is plotted against
number of the atoms trapped in each of the cavities. It is observed
that the nearly perfect state transmission with the fidelity beyond
$0.96$ can be realized when the number of atoms is larger than $20$.

For the swap gate $(\sin\theta_1|0,N\rangle_1 +
\cos\theta_1|1,N-1\rangle_1) \otimes (\sin\theta_2|0,N\rangle_2 +
\cos\theta_2|1,N-1\rangle_2)\rightarrow(\sin\theta_2|0,N\rangle_1 +
\cos\theta_2|1,N-1\rangle_1) \otimes (\sin\theta_1|0,N\rangle_2 +
\cos\theta_1|1,N-1\rangle_2)$, we assume that the "dipole blockade"
effects \cite{LF+01} take place. This allows to ignore the
double-excitation Dicke states in the calculation. The master
equation (\ref{eq:master}) is numerically solved in the space
spanned by the basis vectors (\ref{eq:swapbasis}),
(\ref{eq:swapbasis1}) and $|000\rangle_f|0,N\rangle_1|0,N\rangle_2$.
In the calculation, $g_1=g_2=g$ and $\kappa = \gamma = \beta =0.1g$
are chosen. The average fidelity of the swap gate is defined as
\begin{equation}
F_S =\frac{1}{4\pi^2}\int_0^{2\pi}d\theta_1\int_0^{2\pi}d\theta_2
\vert\langle \Psi_s |\rho(t)|\Psi_s \rangle,
\end{equation}
where $|\Psi_s\rangle=(\sin\theta_2|0,N\rangle_1 +
\cos\theta_2|1,N-1\rangle_1) \otimes (\sin\theta_1|0,N\rangle_2 +
\cos\theta_1|1,N-1\rangle_2)\otimes|000\rangle_f$. For $N=100$, we
obtain the maximal fidelity $0.958$. For $N=10^4$, we obtain the
maximal fidelity $0.992$. In the single-atom scheme \cite{SM+06},
one has to take $\kappa = \gamma = \beta =0.001g$ for obtaining the
same maximum values of the fidelity. Therefore, in the multiatom
scheme, the relatively reliable swap gate can be realized even if
the strength of the coherent interaction between the atoms and the
cavity field is not much larger than the rates of spontaneous
emission and photon leakage.

As shown in section IV, when $\nu \gg \sqrt{N}g$, we may introduce
the resonant mode $c$. Since the fiber mode is not involved in the
mode $c$ according to the transformations \eqref{eq:a2c}, the entangling and
controlled-Z gates are unaffected by fiber losses in this
limitation. In this case, therefore, the master equation of motion
for the atoms and the resonant mode $c$ can be written as
\begin{equation}
  \label{eq:master1}
  \dot{\rho} = -i[H,\rho] + \gamma L[c]\rho + \kappa\sum_{j=1}^2
  \sum_{i=1}^N \Big[2 \sigma_{i}^-(j) \rho \sigma_{i}^+(j) -
  \sigma_{i}^+(j) \sigma_{i}^-(j)\rho - \rho\sigma_{i}^-(j)\sigma_{i}^+(j)\Big],
\end{equation}

For the entangling gate, Eq. \eqref{eq:master1} is solved in the space
spanned by the basis vectors (\ref{basis1}) and
$|0\rangle_c|0,N\rangle_1|0,N\rangle_2$ with the initial state
$|0\rangle_c|1,N-1\rangle_1|0,N\rangle_2$. In the calculation, $g_1=g_2,
\kappa=\gamma=0.1g$ and $\nu =50 \sqrt{N} g$ are chosen. For
concretion, we consider the target state $|\psi_{E1}\rangle=
|0\rangle_c[ |1,N-1\rangle_1 |0,N\rangle_2 + |0,N\rangle_1
|1,N-1\rangle_2] / \sqrt{2}$. Fidelity of the entangling gate is
defined as $F_{E1}=\langle\psi_{E1}|\rho(t)|\psi_{E1}\rangle$. In Fig.
\ref{fig:diss}(b), the
maximum of the fidelity is plotted against the number of atoms. It
is observed that the entangling gate with the fidelity bigger than
$0.95$ is realized when $N\geq 50$.

\begin{figure}[htbp]
  \centering
\includegraphics[width=8cm]{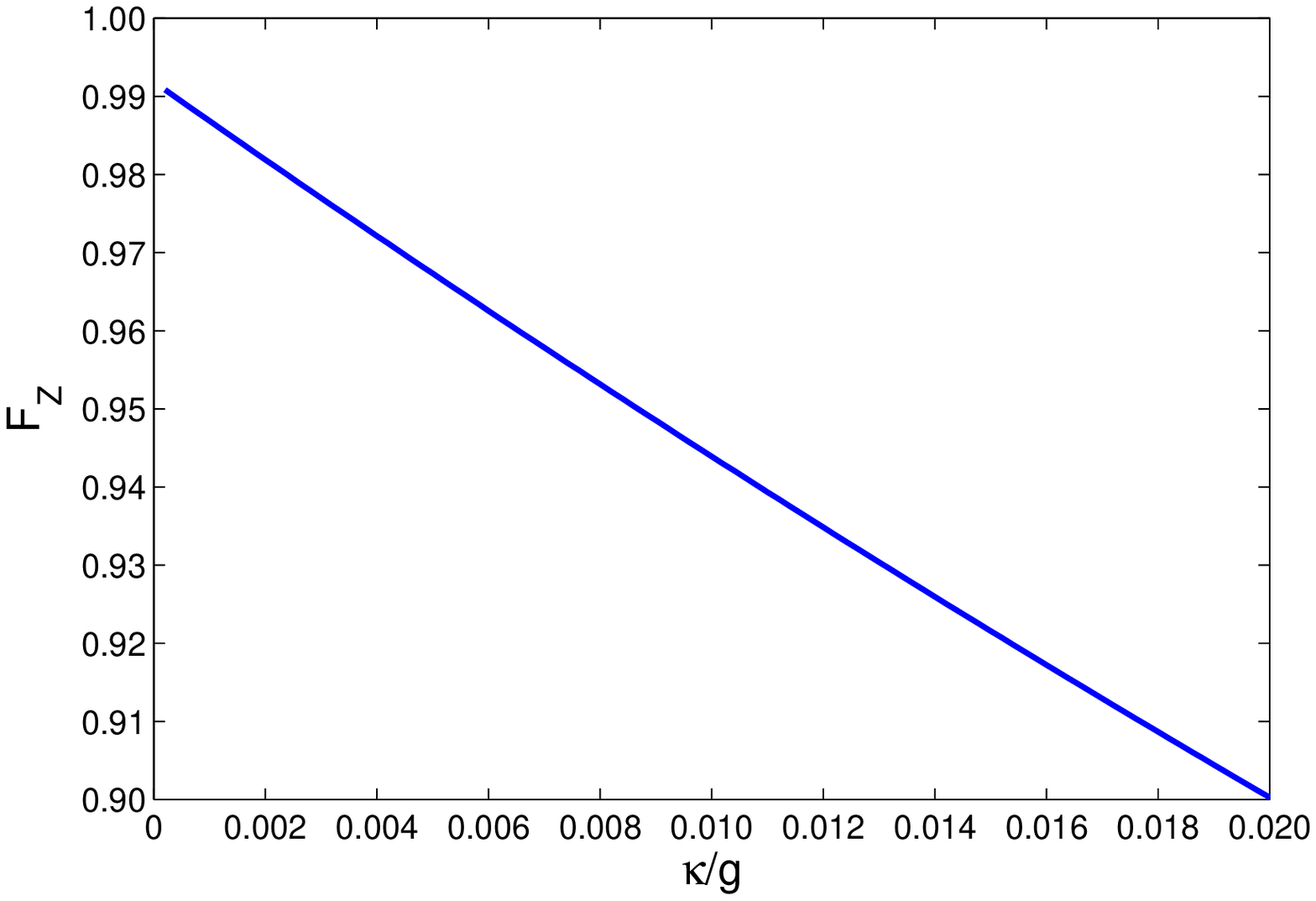}
\includegraphics[width=8cm]{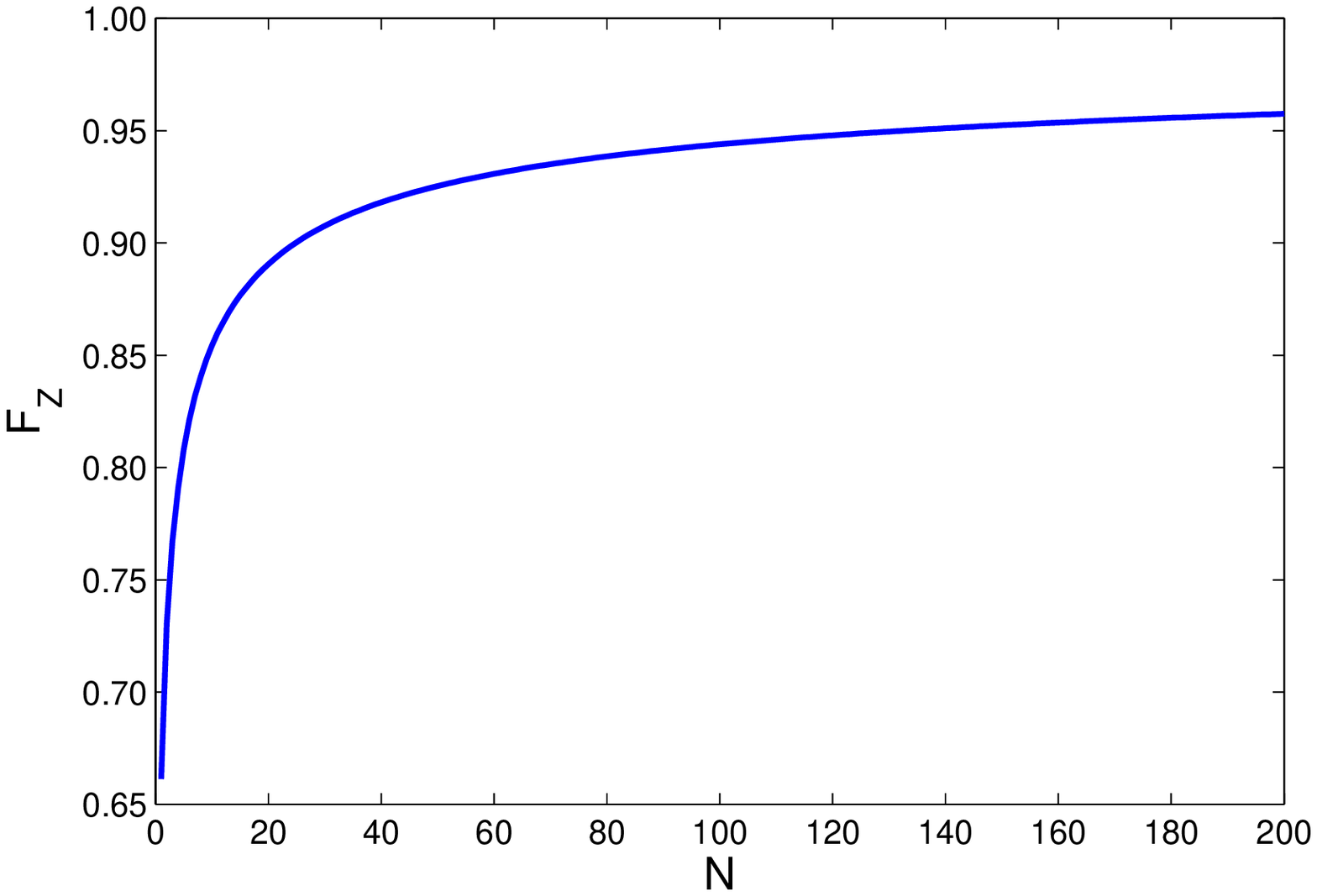}
\caption{(a) Fidelity of the controlled-Z versus the decay rate
$\kappa=\gamma$. (b) Fidelity of the controlled-Z gate as a function
of the atomic number.}
  \label{fig:dissCZ}
\end{figure}

For the controlled-Z gate, Eq. (\ref{eq:master1}) is solved in the space
spanned by the basis vectors
\begin{equation}
  \begin{aligned}
|\psi_{n_c m_1m_2} \rangle&= |n_c\rangle_c |m_1, N-m_1\rangle_1
|m_2,
N-m_2\rangle_2 \\
 \text{with} ~~0\leq n_c,m_1,m_2 &\leq \mathcal{N},
 ~~ n_c+ m_1+m_2 = \mathcal{N},~~ \text{and}
~~ 0\leq m_1, m_2 \leq 1,
  \end{aligned}
\end{equation}
where $\mathcal{N}(=0,1,2)$ is the total excitation number. In the
calculation, we assume that at the initial time the entire system is
in the state $|0\rangle_c\otimes (|0,N\rangle_1 + |1,N-1\rangle_1)\otimes
(|0,N\rangle_2 + |1,N-1\rangle_2)/2$ and take $g_1=g, \delta
= g_2-g_1 = 1.2g,$ and $\gamma=\kappa$. The target state is
$|\Psi_Z\rangle = |0\rangle_c\otimes(|0,N\rangle_1 |0,N\rangle_2 +
|0,N\rangle_1 |1,N-1\rangle_2 + |1,N-1\rangle_1 |0,N\rangle_2 -
|1,N-1\rangle_1 |1,N-1\rangle_2)/2$. Fidelity of the controlled-Z
gate is defined as
$F_Z(t)=\langle\Psi_Z\vert\rho(t)\vert\Psi_Z\rangle$. In Fig.
\ref{fig:dissCZ}(a), the maximum of the fidelity is shown against
the decay rate $\kappa$ when $N=1$. It is observed that the
controlled-Z gate with the fidelity larger than $0.96$ is realized
for $\kappa=\gamma \simeq 0.006 g$. For the controlled-Z gate with
the same fidelity, the condition $\kappa=\gamma \simeq 0.0001g$ is
required in the nonresonant case that was considered in
\cite{SM+06}. Therefore, the resonant scheme under consideration can
highly depress the effects of spontaneous emission and photon
leakage out of the cavity. Now let's consider the multiatom case.
In this case, we take  $\kappa=\gamma=0.1g$ and other parameters
same as above. In Fig. \ref{fig:dissCZ}(b), the maximum of the
fidelity is shown as the number of atoms. It is seen that even for
such a large decay rate the controlled-Z gate with the fidelity
bigger than $0.95$ is achieved when $N\geq 150$.

In the recent experiment \cite{BM+04}, $(34 \mathrm{MHz}, 2.6
\mathrm{MHz}~ \text{and}~ 4.1 \mathrm{MHz})$ for
$(g,\kappa,\gamma)/2\pi$ are achieved. Even more strong-coupling
($\kappa/g \geq 165$) \cite{SKV+05,TH+05} and ultrahigh-Q ($\simeq
10^8$) microcavity\cite{SKV+05} have been predicted and proved
experimentally. Therefore, the condition $g\sim10\gamma$ that is
required in the present scheme can be satisfied by use of the
developed techniques. As for the fiber coupling to the cavities, a
perfect fiber-cavity coupling (with efficiency larger than $99.9\%$)
can be realized by fiber-taper coupling to high-Q silica
microspheres \cite{SK+03}. On the other hand, it is also possible to
locate atoms in an optical cavity with the spatial precision
$\lambda/10$ \cite{MP+05}. Therefore, we should believe that it is
feasible with present techniques to realize the quantum processes
under consideration in this paper.

\section{Summary}

We investigate how to deterministically implement quantum state
transfer, and swap, entangling and controlled-Z gates between two
subsystems consisting of two-level atoms trapped in single-mode
cavities spatially separated and connected by an optical fiber. If
the atoms collectively and resonantly interact with the local
fields, we find that a perfect quantum state transfer can be
realized if the ratio of the coupling constant between the atoms and
the cavity field to the coupling constant between the fiber and
cavity modes satisfies the established condition. If the event for
two atoms to be simultaneously excited in the same cavity is
depressed by the "dipole blockade" effect, it is shown that a nearly
perfect swap gate can realized. When the coupling between the fiber
and cavity modes is much stronger than the coupling between the
atoms and the cavity field, the entire system can be approximated as
two qubits resonantly coupling to a harmonic oscillator. In this
limitation, the nearly perfect entangling and controlled-Z gates can
be realized. It is also noticed that the quantum computation
precesses under investigation are much robust against the variation
of the coupling constants. We find that the operation time of the
quantum state transfer and the quantum logic gates is proportional
to $1/\sqrt{N}$ where $N$ is number of the atoms in each of the
cavities. Therefore, the quantum computation processes can be
greatly speeded up when the number of atoms becomes large. This
effect is useful for depressing the influence of unavoidable
decoherence processes such as spontaneous emission and photon
leakage. By numerically solving the master equation of motion for
the entire system, we investigate the influence of spontaneous
emission of the atoms and photon leakage out the cavities and fiber
on the quantum computation processes. In the calculation, we assume
that the atoms collectively interact with the privileged cavity
modes but individually decay to other modes of the electromagnetic
field. We find that when the number of atoms is large the quantum
computation processes can be accomplished with high fidelity even if
rates of the dissipations are large around the order of one-tenth of
the coherent coupling strength between the atoms and local cavity
fields.

The above conclusions are based on the following assumptions: (1)
there does not exist direct interaction between the atoms when all
the atoms are in the ground state; (2) the atoms collectively
interact with the cavity mode; (3) the dipole blockade effect takes
place; (4) the atomic spontaneous emission is individual. We now
investigate if these conditions can be fulfilled in current
experiments. For a concrete purpose, let us take alkali atomic
levels $7s_{1/2}$ and $np_{3/2}(n\sim50)$ as the ground state
$|g\rangle$ and the excited state $|e\rangle$ in the present
discussion, respectively \cite{VV+06}. Since the atom resonantly
interacts with the cavity field, corresponding the atomic
transition, the wavelength of the cavity field ranges from $700$ to
$800$ nm. Suppose that the cavity field has the spatial distribution
$E(r,z)=E_0e^{-\frac{r^2}{W^2}}\cos(kz)$£¬where $W$ is the waist
width and $k$ is the wave vector of the cavity field \cite{HL+00}.
The coupling strength between the cavity field and the atom is given
by $g(r,z)=g_0e^{-\frac{r^2}{W^2}}\cos(kz)$. Suppose that all the
atoms are trapped in a shaped-flat-disk trap of transverse radius
$\delta r$ and axis-thickness $\delta z$ in the cavity. This
flat-disk trap with the radius $\delta r=2.3$ $\mu$m and the thickness
$\delta z=33$ nm has been realized in experiments \cite{BM+04}. The
no interaction condition requires $d \gg r_{g}$, where $d$ is the
average distance between two atoms and  $r_g$ is the radius of the
atom in the ground state. Since $\delta r \gg \delta z$, we may
approximately consider that the atoms have a surface distribution in
the trap. In this way, the average distance may be estimated by
$d=\sqrt{\pi(\delta r)^2/N}$ , where N is the number of atoms in the
trap. Consider an alkali atom with the valence electron in $ns$
state. The orbit radius of the valence electron is approximately
given by $n^2a_0$, where $a_0$  is the Bohr radius. With current
experiment parameters \cite{BM+04}, for $7s_{1/2}$ of Cs atom, we
have $r_g\approx 2.6$ $nm$ and $d\approx 4.1N^{-1/2}$ $\mu$m. Therefore,
the no direct interaction condition can be satisfied very well when
$N<200$. The collective interaction condition requires that the
variation of the coupling strength in the trap must be very small,
that is, $\delta g/g_0 \ll 1$ , or $k\delta z \ll 1$ and $\delta r
\ll W$. In the current experiment \cite{BM+04}, $k\delta
z\approx0.26$ and $\delta r\approx0.1W$. The spatial variation in
the trap results in the change of the coupling strength $\delta
g/g_0 < 10^{-2}$. Thus, all the atoms in the trap have the nearly
same coupling strength and then collectively interact with the
cavity field. We now check the dipole blockade condition. The dipole
blockade mechanism assumes that a strong interaction such as either
the van der Walls interaction \cite{TF+04,SR+04,LR+05} or
dipole-dipole interaction \cite{VV+06} takes place when two atoms in
an alkali atomic ensemble are simultaneously excited into Rydberg
states. This strong interaction can make the Rydberg level shifts so
large that all the transitions into states with more than a single
excitation are blocked \cite{LF+01}. This excitation blockade
phenomenon has been demonstrated in several experiments
\cite{TF+04,SR+04,LR+05,VV+06}. In order to guarantee the occurrence
of the dipole-blockade mechanism, the condition $d\sim r_e$ is
required, where $r_e$ is the radius of the atom in the excited
state. For the Rydberg level $50p_{3/2}$ \cite{VV+06}, $r_e\approx
0.13$ $\mu$ m. For the Rydberg level $80p_{3/2}$
\cite{TF+04,SR+04,LR+05}, $r_e \approx 0.34$ $\mu$m. For $N=100$,
$d\sim 0.4$ $\mu$m. Thus, the dipole-blockade mechanism can take place
when $N>100$. From the above discussion, we see that the no
interaction condition has to be traded off to the dipole blockade
condition. In this way, the number of atoms in the trap is limited
around $200$ with the current experiment parameters. Therefore, with
the current cavity and trap techniques \cite{MP+05,NM+05}, the
present multiatom scheme may speed up the operation of the logical
gates by $10$ times. As regards spontaneous emission, for $N=100$,
from the above parameters, we have $kd\sim3$, that is, along the
transverse direction, the average distance of atoms is larger than
the reduced wavelength of the cavity field. On the other hand, only
one atom is excited if the dipole blockade effect takes place.
Therefore, in the present scheme, we can consider that the excited
atoms individually couple to all transverse modes of the
electromagnetic field \cite{LF+01, DLCZ01,FT02}. Here, the key point
is that the atoms are trapped in the flat-disk-trap. The trap is so
narrow along the cavity axis and slowly varying in the transverse
direction that the atoms collectively interact with the cavity field
but so wide in the transverse direction that the atoms individually
decay. In summary, according to parameters in current experiments,
the essential assumptions employed in the present scheme are
reasonable and realistic. The multiatom quantum state transfer and
quantum logic gates could be realized with current optical cavity
and atomic trap techniques.

\acknowledgments Zhang-qi Yin thanks Hong-rong Li, Peng Peng and
Ai-ping Fang for valuable discussions. This work was supported by
the Natural Science Foundation of China (Grant Nos. 10674106,
10574103, 05-06-01).


\end{document}